\documentclass[twocolumn,english,final]{article}
\pdfoutput=1

\usepackage[utf8]{inputenc}
\usepackage[T1]{fontenc}

\usepackage{cite}

\usepackage{abstract}

\usepackage[fleqn]{amsmath}
	\allowdisplaybreaks[1]
\usepackage{amsfonts}
\usepackage{amssymb}

\usepackage{microtype}

\usepackage[toc]{appendix}

\usepackage{hyperref}
	\hypersetup{%
	pdfauthor={Federico Di Gioia and Giulia Maniccia and Giovanni Montani and Jacopo Niedda},%
	pdftitle={nonunitarity Problem in Quantum Gravity Corrections to Quantum Field Theory with Born-Oppenheimer Approximation}%
	}

\newcommand{\eu}{\mathrm{e}}
\newcommand{\iu}{\mathrm{i}}
\newcommand{\di}{{}\, \mathrm{d}}

\DeclareMathOperator{\const}{const}

\newcommand{\spR}{\ensuremath{{}^{(3)} \! R}}

\newcommand{\D}{\mathrm{D}}
\newcommand{\Db}{\bar{\mathrm{D}}}

\newcommand{\realPart}{\ensuremath{\mathbb{R}\mathrm{e}}}
\newcommand{\imaginaryPart}{\ensuremath{\mathbb{I}\mathrm{m}}}

\newcommand{\ord}[1]{\mathcal{O} \left(#1\right)}

\usepackage{mathtools}

\DeclarePairedDelimiter{\avg}{\langle}{\rangle}
\newcommand{\avgt}[1]{\widetilde{\avg{#1}}}

\begin{document}
\title{Nonunitarity problem in quantum gravity corrections to quantum field theory with Born-Oppenheimer approximation}

\author{%
	Federico Di Gioia\textsuperscript{1}\thanks{\href{mailto:federico.digioia@uniroma1.it}{federico.digioia@uniroma1.it}},
	Giulia Maniccia\textsuperscript{1}\thanks{\href{mailto:giulia.maniccia@roma1.infn.it}{giulia.maniccia@roma1.infn.it}}, Giovanni Montani\textsuperscript{1,2}\thanks{\href{mailto:giovanni.montani@enea.it}{giovanni.montani@enea.it}}, Jacopo Niedda\textsuperscript{1}
\thanks{\href{mailto:jacopo.niedda@uniroma1.it}{jacopo.niedda@uniroma1.it}} \\%
	\small {}\textsuperscript{1}Physics Department, ``La Sapienza'' University of Rome, P.le A. Moro 5, 00185 Roma, Italy\\%
	\small {}\textsuperscript{2}ENEA, C.R. Frascati (Rome), Italy Via E.\ Fermi 45, 00044 Frascati (Roma), Italy%
}
\date{}
	
\twocolumn[
		\maketitle
		
		\begin{onecolabstract}
		The problem of time is one of the most relevant open issues in canonical quantum gravity. Although there is a huge literature about this topic, a commonly accepted solution has not been found yet. Here, we focus on the semiclassical approach to the problem of time, that has the main goal of reproducing quantum field theory on a fixed Wentzel-Kramers-Brillouin (WKB) background accounting also for quantum gravity corrections. We analyze the different choices of the expansion parameter and discuss the problems arising in previous proposals, where a nonunitary evolution emerges as an effect of quantum gravity corrections. In this work, we develop a new approach to solve this problem by performing the WKB expansion with the introduction of the so-called kinematical action as a clock for quantum matter, that allows to recover a unitary dynamics.
		\vspace*{\baselineskip}
		\end{onecolabstract}
]

\saythanks

\section{Introduction\label{sec:Intro}}
	The application of canonical quantization procedures to general relativity (GR) is the most traditional attempt to derive a quantum theory of the gravitational field. Canonical quantization --- especially of gauge field theories --- has mathematical ambiguities, but can at least yield an approximation of the real quantum theory and a good framework to deal with its main issues~\cite{bib:dirac-lecturesOnQuantumMechanics}.
	
	One of the long-standing problems of canonical quantum gravity is the so-called \emph{frozen formalism}, i.e.\ the absence of an evolution of the quantum gravitational field with respect to an external clock~\cite{bib:cianfrani-canonicalQuantumGravity}.
	
	Over the years, many approaches have been proposed to address this question, based both on introducing time through some matter source~\cite{bib:kuchar-1991,bib:brown-1995} or identifying it with an internal source-time variable~\cite{bib:isham-1992}; they can differ among each other also for constructing time variables before or after the canonical quantization procedure has been performed. The common starting point of these works is the concept of \emph{relational time}~\cite{bib:rovelli-1991}: by imposing suitable boundary conditions, a chosen subsystem can be properly adopted as a clock for the remaining part of the quantum system. However, these approaches seem to be in contradiction with the fundamentals of quantum mechanics that time is an external parameter and measurements are performed by a classical observer. On the basis of a relational time approach, it is not clear how to reproduce the proper limit of quantum field theory on a curved background, starting from the Wheeler-DeWitt (WDW) equation~\cite{bib:deWitt-1967}.
	
	In this respect, a different proposal has been investigated in~\cite{bib:vilenkin-1989}, where the situation considered is such that the quantum system can be separated into a set of semiclassical WKB variables and a ``small'', fast, purely quantum component. This scenario represents the quantum gravity version of a Born-Oppenheimer (BO) approximation, with the peculiar feature that now the dependence of the fast quantum system on the semiclassical variables can be used to reintroduce the notion of an external time for the fast system component, essentially coinciding with the standard label time of the space-time slicing. This approach, which can be applied to any set of variables (see, e.g.,~\cite{bib:montani-2009}), is particularly appropriate to reconstruct the limit of quantum field theory on a classical curved background from quantum gravity. For a discussion on the necessary conditions to adopt the BO approximation and applications, see~\cite{bib:montani-2017}.
	
	The study~\cite{bib:vilenkin-1989}, which represents the starting point in this path, is based on the WKB expansion of the gravity-matter system, performed using the Planck constant as the natural expansion parameter and cutting the dynamics up to first order in~$\hbar$. A similar proposal is implemented in~\cite{bib:kiefer-1991}, but using an expansion parameter $M$ proportional to the square of the Planck mass (\emph{de facto} the Newton constant) and expanding the dynamics up to, in principle, any order of approximation. This study has the merit to arrive to similar results than the former~\cite{bib:vilenkin-1989}, but without requiring the rapid variation of the system wave function with respect to the small, quantum subsystem variables.
	
	The emerging problem in~\cite{bib:kiefer-1991} is that, as far as the next order of approximation is considered, which corresponds to quantum gravity corrections to quantum field theory, a nonunitary character of the quantum dynamics emerges. This result prevents the predictivity of the approach at such order. Nonetheless in~\cite{bib:kiefer-2016} the study of cosmological perturbations on a classical isotropic Robertson-Walker background is developed in the framework of the quantum gravity corrections found above: the application shows the modification of the inflationary spectrum of perturbations, due to non-classical effects of the gravitational field, and the smallness of the nonunitary contributions. 
	
	Subsequently, two different proposals to solve the nonunitary problem of the WKB theory, at the order of quantum gravity corrections, have been developed in~\cite{bib:bertoni-1996} and~\cite{bib:kiefer-2018}. The proposed solutions rely on two different points of view: one aims to define a conserved probability density, disregarding the details of the evolution quantum operator, with an extended (gauge invariant) BO approximation; the other aims to reconstruct \emph{a posteriori} a well-behaving Schr\"odinger evolution of the quantum subsystem, by altering the pure WKB dynamics of the gravitational background.
	
	
	In this study, we offer a critical analysis of these proposals and outline how the fundamental problem of dealing with nonunitary contributions in the quantum dynamics has not been properly addressed yet. In fact, regarding the proposal~\cite{bib:bertoni-1996}, apart from completing the analysis by properly rescaling the background wave function (which implies an important cancellation of the backreaction that quantum matter exerts on the background), we clarify how the evolution operator still presents the nonunitary features outlined in~\cite{bib:kiefer-1991} and we stress that no real Hilbert space is constructed in this approach, since the scalar product of two different states is not dynamically preserved.
	
	On the other hand, the approach~\cite{bib:kiefer-2018},	based on using the Hamiltonian and corrected Hamiltonian complex eigenvalues instead of the operators, manages to remove the nonunitary terms via a phase redefinition of the quantum wave function, using two problematic requirements: (i) the spectrum of the time derivative of the corrected Hamiltonian operator is constructed with the time derivatives of the respective eigenvalues and (ii) the matter Hamiltonian operator and its time derivative must commute. These two features are here shown to be not valid in general and therefore this procedure must be considered an \emph{ad hoc} solution valid in specific situations only. A more subtle problem of this approach is the lack of gauge invariance under the phase transformations performed on the wave functions, a feature that is instead present in~\cite{bib:bertoni-1996}.
	
	We here develop a proposal to solve	the nonunitary problem discussed above, based on a different construction of the physical clock for the gravity-matter system in the considered WKB separation of the dynamics. The main concept of this proposal is the introduction of the so-called kinematical action~\cite{bib:kuchar-1981}, see also~\cite{bib:montani-2002}, as a clock for quantum matter, which modifies the results of the WKB expansion performed in the parameter $M$ related to the Planck mass (as in~\cite{bib:kiefer-1991}). This component consists of an additional Hamiltonian term that reinstates the covariance of the theory under Arnowitt-Deser-Misner (ADM) foliation, thus recovering a covariant construction of the parabolic constraints for quantum matter fields on a curved classical background. In fact, by fixing the background metric in a given reference frame, the lapse function and the shift vector would be naturally fixed without the introduction of such term, so that the dynamics of a quantum field on that space-time representation would no longer be associated to the Dirac constraints from which a functional Schr\"odinger dynamics would naturally emerge.
	
	As a result of the implementation of the \emph{kinematical action}, the quantum dynamics of the field is characterized by parabolic constraints, linear in the momentum canonically conjugate to the	four-dimensional variables, thought as fields depending on the slicing space-time variables.	Then, making use of the expression of the	deformation vector, the resulting Dirac implementation of these constraints turns out to coincide with a functional Schr\"odinger equation.
	
	Here, the kinematical action, which is in principle added to the full quantum system of gravity and matter, is regarded as a fast quantum component, on the same footing of the real quantum matter field. Thus, in the present context, the fluid associated to the kinematical action does not appear in the Hamilton-Jacobi equation and the standard Einsteinian dynamics of the gravitational background is unaffected, as viewed at the order $M$ of the considered perturbative expansion.
	
	At the order zero in the parameter $M$, we are able to recover a standard functional Schr\"odinger equation for the quantum field, overlapping to standard quantum field theory. Finally, at the order of expansion $1/M$, we arrive to write down a Schr\"odinger equation containing corrections from the quantum nature of the gravitational field, as viewed in a WKB expansion of the associated vacuum Wheeler-DeWitt equation. The modification is linked to the classical function which is solution of the Hamilton-Jacobi equation and it is responsible for a term, whose morphology is clearly shown to be unitary.
	
	The results of the approach here presented offer a new investigation tool to evaluate the effect of a nonpurely classical dynamics of quantum field theory, in the limit of very small energies involved in the quantum dynamics with respect to the Planckian scale.
	
	The structure of the paper is as follows. In Sec.~\ref{sec:WKB} we present the WKB approach to canonical quantum gravity, starting in Sec.~\ref{ssec:BasicFormalism} with the basic formalism of canonical quantum gravity. In Sec.~\ref{ssec:expansionVilKief} the two WKB semiclassical expansions~\cite{bib:vilenkin-1989,bib:kiefer-1991} (i.e.\ in $\hbar$ and in $M$ related to the Planck mass, respectively) are critically reviewed, with a comparison between the two approaches and extension of the $\hbar$ expansion to arbitrary orders. In Sec.~\ref{ssec:non-unit_Plmassexp}, we critically analyse the procedure used in~\cite{bib:kiefer-2018} to cancel the non-Hermitian terms of the Hamiltonian operator and obtain a unitary dynamics, arguing that the used assumptions are not true in general.
	
	In Sec.~\ref{sec:BOExpansion} we focus on the expansion based on the exact decomposition of the wave function with BO approximation; the procedure of~\cite{bib:bertoni-1996} is summarized in Sec.~\ref{ssec:BO-Venturi} and then completed in order not to break the gauge invariance of the approach. In Sec.~\ref{ssec:EnhancementBO} we show the issues still present within this procedure, concluding that it does not resolve the nonunitarity problem, due to the requirements which are not satisfied in the general case and the Hilbert space which is truthfully constructed only with the scalar product between the same states.
		
	Section~\ref{sec:AzioneCin} contains the proposal to bypass the problem of nonunitarity by including the kinematical action. First we show in Sec.~\ref{ssec:OriginNonUn} the origin of nonunitary correction terms arising in the previous proposals, due to the definition of a WKB time parameter in terms of the dependence of the quantum subsystem on the classical background variables. In Sec.~\ref{ssec:IntrodAzioneCin} we introduce the kinematical action as presented in~\cite{bib:kuchar-1981}, showing that it restores the covariance of the theory under choice of ADM splitting for a fixed gravitational background; in Sec.~\ref{ssec:UnitaryEv} we then insert the kinematical action in the theory of quantum matter on a canonically quantized gravitational background and perform the WKB expansion in powers of the same parameter $M$, related to the Planck mass, used in~\cite{bib:kiefer-1991}. The results of such expansion show that, with the definition of time through the kinematical action variables, the classical limit for gravity and the standard Schr\"odinger evolution for quantum matter are correctly reproduced at the orders $M$ and $M^0$; the expansion at order $1/M$ brings the corrections to quantum matter dynamics caused by the quantum nature of the gravitational background arising from the Wheeler-DeWitt equation, which are shown in this paper to be unitary, thus solving the problem emerging in the previous proposals. 
	
	In Sec.~\ref{sec:Conclusions} the concluding remarks are presented.	Appendixes A and B contain discussions on some technical aspects.
	
\section{The WKB Expansion of Quantum Gravity \label{sec:WKB}}
	After an introduction of the basic formalism of canonical quantum gravity, we here analyze the two semiclassical expansions~\cite{bib:vilenkin-1989}~up to the quantum mechanical order, and~\cite{bib:kiefer-1991}~up to the first quantum gravity corrections, expanded in~\cite{bib:kiefer-2018} to any desired order. We will unify these approaches in a single expansion, showing that they both lead to nonunitary dynamics.
	
	\subsection{Frozen formalism in canonical quantum gravity\label{ssec:BasicFormalism}}
		Before applying the canonical quantization procedure, we shall briefly recall the Hamiltonian formulation of GR. In the general case, such approach leads to the concept of \emph{superspace}, i.e.\ the configuration space of all the geometric and matter variables. It naturally follows that, since the variables are fields defined over a curved space-time, the full theory has a functional nature and requires some renormalization procedure to yield finite predictions.
		
		In some cases, when highly symmetric space-times are involved, it is possible to reduce the dynamics to a finite-dimensional scheme and replace the concept of superspace by its finite-dimensional analogous, i.e.\ \emph{minisuperspace}: this procedure will be applied in the next section, as it is implemented in the work~\cite{bib:vilenkin-1989}, and implicitly assumed in~\cite{bib:kiefer-1991}. The reduced theory finds its main applications in cosmology, where homogeneous space-times are considered, such as the previously mentioned~\cite{bib:kiefer-2016}.
		
		Let us consider a universe filled by matter fields. We write the universe wave function as
		\begin{equation}
		\label{eq:defPsi}
			\Psi = \Psi (\{h_{ij}\}, \phi_a) ,
		\end{equation}
		which depends on the equivalence class of 3-geometries~$ h_{ij} $ and on the matter fields~$ \phi_a $. Following~\cite{bib:arnowitt-1962}, we start from the WDW equation:
		\begin{subequations}
		\begin{gather}
		\label{eq:WDW}
			H \Psi = \mathcal{H}_\mathrm{g} \Psi + \mathcal{H}_\mathrm{m} \Psi = 0\\
		\label{eq:Hg}
			\begin{split}
				\mathcal{H}_\mathrm{g} = &-\frac{2 \hbar^2 \kappa}{\sqrt{h}} G_{ijkl} \frac{\delta^2}{\delta h_{ij} (x) \, \delta h_{kl} (x)}\\
				&- \frac{\sqrt{h} \spR}{2 \kappa}
			\end{split}\\
		\label{eq:Hm}
			\begin{split}
				\mathcal{H}_\mathrm{m} = &-\frac{\hbar^2}{2 \sqrt{h}} G_{ab} \frac{\delta^2}{\delta \phi_a (x) \, \delta \phi_b (x)}\\
				&+ u(h_{ij}, \phi_a) ,
			\end{split}
		\end{gather}
		\end{subequations}
		where $\kappa = 8\pi G/c^3$ and \eqref{eq:Hg}, \eqref{eq:Hm} are the super-Hamiltonian functions of the gravitational field and matter fields respectively. We have assumed for simplicity the matter component to consist in a set of self-interacting scalar fields~$ \phi_a $, minimally coupled with the geometry, and with total potential energy
		\begin{equation}
			u(h_{ij}, \phi_a) = \sum_a u_a (h_{ij}, \phi) .
		\end{equation}
		However, the following analysis is more general and the results hold for any choice of the matter component. We will use natural units for the speed of light:~$ c \equiv 1 $.
		
		The tensor
		\begin{equation}
			G_{ijkl} = \frac{1}{2} \left( h_{ik} h_{jl} + h_{il} h_{jk} - h_{ij} h_{kl} \right)
		\end{equation}
		is the supermetric of the geometric subspace, while
		\begin{equation}
			G_{ab} = \delta_{ab}
		\end{equation}
		is the supermetric of the matter subspace, $ \delta_{ab} $ being the Kronecker delta.
	
		The spatial curvature $ \spR $ defines a geometric superpotential
		\begin{equation}
			V = - 2 \sqrt{h} \spR .
		\end{equation}
		which can be modified through $ \spR \rightarrow \spR - 2 \Lambda $ to include a cosmological constant term. Once the cosmological constant is included, the model can be used to describe the inflationary phase of the early universe.
		
		We stress that, in the general nonhomogeneous case, there are additional constraints to be satisfied by the wave function, containing the so-called supermomentum functions of the geometric and matter sectors. Since they are not present in the works analyzed in Secs.~\ref{sec:WKB} and~\ref{sec:BOExpansion}, these terms will be properly addressed and implemented in Sec.~\ref{sec:AzioneCin}.
		
		The dynamics of the model is fully encoded in the super-Hamiltonian constraint~\eqref{eq:WDW}, which results in the well-known frozen formalism problem: by applying the canonical quantization procedure to the WDW equation~\eqref{eq:WDW}, it is immediately found that the universe wave function does not have an explicit dependence on time
		\begin{equation}
		\label{eq:problemtime}
			\iu\hbar \frac{\partial}{\partial t} \Psi = \hat{H} \Psi = 0
		\end{equation} 
		which is rather disturbing, and has encouraged physicists to look for a new definition of time parameter, i.e.\ a relational time~\cite{bib:rovelli-1991} that describes a proper, nontrivial evolution of the universe. In fact, the WDW equation being satisfied by time-independent quantum states does not mean that the universe is static: the condition is a direct consequence of the fact that GR naturally has a parametrized Hamiltonian formulation, due to time reparametrization invariance. However, an external time label is an essential ingredient for a quantum dynamical theory. Hence, one should look for a meaningful definition of time among the minisuperspace matter and geometric variables. A suitable construction of the time parameter can be performed by a semiclassical separation and expansion of the components of the universe, resulting in promising effective theories; a solution to the complete problem of quantum gravity, i.e.\ a procedure valid at any scale, has not been found yet.
		
		In this section, to simplify our discussion, we choose normal ordering by placing the minisupermetric components on the left of the derivative operators. Moreover \cite{bib:kiefer-1991} showed that, at least for the WKB expansion we are going to perform here, such ordering ambiguities are absorbed in the classical part of the expansion and do not influence the quantum Schr\"odinger equation. The more general ordering of the operators will be restored in Sec.~\ref{sec:AzioneCin}.
		
		In order to write the WDW equation in a more compact and clear form, following~\cite{bib:vilenkin-1989}, we decompose the variables in two subsets: the classical and the quantum ones. We refer generally as $ c = \{c_\alpha\} $ to the subset of classical variables and as $ q = \{q_\nu\} $ to the subset of quantum ones. Moreover, we redefine the supermetric as a unique tensor as
		\begin{equation}
			\mathcal{G}_{ab} = \frac{1}{\sqrt{h}} G_{ijkl}, \quad a, b = \{i, j\}, \{k, l\}
		\end{equation}
		for gravitational variables and as
		\begin{equation}
			\mathcal{\tilde{G}}_{ab} = \frac{1}{\sqrt{h}} G_{ab}
		\end{equation}
		for matter ones, possibly including any necessary constant. This way, we can write the Laplacian operators in a compact way:
		\begin{subequations}
		\begin{gather}
		\label{eq:defNablac}
			\nabla_c^2 \equiv \mathcal{G}_{ab} \frac{\delta^2}{\delta h_a \, \delta h_b}\\
		\label{eq:defNablaq}
			\nabla_q^2 \equiv \mathcal{\tilde{G}}_{ab} \frac{\delta^2}{\delta \phi_a \, \delta \phi_b} .
		\end{gather}
		\end{subequations}
		Then, the total Hamiltonian~\eqref{eq:WDW} reads
		\begin{equation}
		\label{eq:WDWcompact}
			H \Psi = \left( - K \nabla_c^2 + U_c + H_q \right) \Psi = 0 ,
		\end{equation}
		where $ K $ is a constant dependent on the choice of the expansion parameter, $ U_c $ is the classical potential and $ H_q $ is the quantum Hamiltonian. We stress that this formalism applies to the choice of ordering that provides general covariance in the minisuperspace~\cite{bib:vilenkin-1989}, when the second order derivative operators $ \nabla_c^2 $ and $ \nabla_q^2 $ are considered no more as Laplacians, but as Laplace-Beltrami operators.
		
	\subsection{WKB expansion in Planck constant and Planck mass \label{ssec:expansionVilKief}}
		The semiclassical WKB approach is an attempt to define a time label in the limit in which some of the variables in the universe can be treated classically.
		
		Classical variables determine a fixed background over which it is possible to define the time evolution of a quantum subsystem. The presence of such variables is needed to define a time label that ensures the positive semidefiniteness of the Klein-Gordon-like scalar product induced by the WDW equation, and finds a conceptual justification in the role played by classical devices in the interpretation of quantum measurement~\cite{bib:vilenkin-1989}.
		
		The core idea of the semiclassical approach is that Eq.~\eqref{eq:WDWcompact} may be solved perturbatively in some quantum parameter, e.g.\ the Planck constant~$ \hbar $~\cite{bib:vilenkin-1989} or some parameter depending on the gravitational constant~$ G $, such as the Planck mass~$ m_\mathrm{P} $~\cite{bib:kiefer-1991}. In both cases the wave function is decomposed in a WKB semiclassical wave function for the background and in a wave function for the quantum subsystem. This procedure was developed by~\cite{bib:vilenkin-1989} with expansion parameter $\hbar$ in a minisuperspace model, and by~\cite{bib:kiefer-1991} with the expansion parameter
		\begin{equation}
		\label{eq:defM}
			M \equiv \frac{1}{4 c^2 \kappa} = \frac{c m_\mathrm{P}^2}{4 \hbar} ,
		\end{equation}
		where $ m_\mathrm{P} $ is the reduced Planck mass. Later, this same procedure was extended by~\cite{bib:kiefer-2018} up to arbitrary orders; for further discussion on this approach see also~\cite{bib:chataignier-2019, bib:chataignier-2020}. We shall underline that, even though it was stated by the author in~\cite{bib:kiefer-1991} that the latter expansion is performed in the generic superspace, the supermomentum constraint is not implemented in that theory. In review of~\cite{bib:kiefer-1991}, such contribution is not added here; it will be reinstated for the sake of generality in Sec~\ref{sec:AzioneCin}, which is set in the superspace.
		
		Besides some structural differences, mathematically these expansions are very similar. Both of them make use of an adiabatic approximation to separate the semiclassical background from the quantum subsystem, resembling a Born-Oppenheimer approximation which, however, is mathematically realized in different ways. Here, we present them in a critical way, while also extending the formalism of~\cite{bib:vilenkin-1989} up to arbitrary orders of expansion, leading to nonunitary corrective terms, and unifying the procedure of~\cite{bib:kiefer-2018} with the~$ \hbar $ expansion.
		
		In the $ \hbar \rightarrow 0 $ expansion, Eq.~\eqref{eq:WDWcompact} reads
		\begin{equation}
			\left( - \hbar^2 \nabla_c^2 + U_c (c) + H_q \right) \Psi (c, q) = 0 ,
		\end{equation}
		with
		\begin{gather}
			H_c = - \hbar^2 \nabla_c^2 + U_c (c)\\
			H_q = - \hbar^2 \nabla_q^2 + U_q (c, q) .
		\end{gather}
		where the operator $ H_c $ is the part of the Hamiltonian obtained neglecting all the quantum variables.
		
		This decomposition in classical and quantum variables requires some assumptions. First, we assert that the quantum Hamiltonian is small with respect to the matter one, expressed as
		\begin{equation}
		\label{eq:smallness}
			\frac{\hat{H}_q \Psi}{\hat{H}_c \Psi} = \ord{\hbar} .
		\end{equation}
		Second, we assume the classical and quantum subspaces to be orthogonal and the supermetric of the classical subspace to depend on classical variables only. This means
		\begin{subequations}
		\begin{gather}
			\mathcal{G}_{\alpha\beta} = \mathcal{G}_{\alpha\beta} (c)\\
			\mathcal{G}_{\alpha\nu} = 0 ,
		\end{gather}
		\end{subequations}
		where $ \alpha, \beta $ are indices over classical variables and $ \nu $ over quantum ones. The last equation is a stronger form of one of the core assumptions of~\cite{bib:vilenkin-1989}, i.e. $ \mathcal{G}_{\alpha\nu} = \ord{\hbar} $: it is needed to extend the expansion after the quantum mechanical order~$ \ord{\hbar} $ studied in the original paper.
		
		On the other hand, in the $ M \rightarrow \infty $ expansion Eq.~\eqref{eq:WDWcompact} becomes
		\begin{equation}
			\left( - \frac{\hbar^2}{2 M} \nabla_g^2 + M V (g) + H_m \right) \Psi (g, m) = 0 .
		\end{equation}
		where the constant $ M \propto 1 / G $ defined in~\eqref{eq:defM} will force the gravitational variables to be classical and the matter ones to be quantum, as will soon be clear. For this reason, we identify the variables as $ c = g $ and $ q = m $. This separation is backed by the strong assumption given by the limit $ M \rightarrow \infty $ corresponding to $ G \rightarrow 0 $, which intuitively implies a vacuum universe. For this reason, \emph{ad hoc} procedures may be needed to take into account a classical matter component, such as a rescaling of the matter fields by $ M $ as performed in~\cite{bib:kiefer-2016}: this is more likely an attempt to work around the problem, since redefining the fields through the expansion parameter is not conceptually satisfying.
		
		The obvious advantage of this choice of parameter is that no further assumptions are needed: this one alone is enough to decompose classical and quantum parts. Since $ M $ has the dimension of a mass over a length, this expansion is expected to hold for particles with small mass over Compton length ratio, which happens for particles whose mass is much smaller than the Planck mass.
		
		The nasty difference in the assumptions has a very simple origin: Eq.~\eqref{eq:WDWcompact} has a perfect symmetry between geometric and matter terms with respect to the order of $ \hbar $, while with respect to $ M $ (i.e.\ $ \kappa $) there is one order gap between them. In the $\hbar$ expansion, this gap is precisely recovered with the additional hypothesis of smallness.
		
		What is missing here with respect to a true BO approximation is the procedure of averaging over the quantum variables, that may allow for the introduction of a backreaction; this feature is instead present in~\cite{bib:bertoni-1996}, which we will deal with in Sec.~\ref{sec:BOExpansion}.
		
		Let us now write the wave function as
		\begin{equation}
			\Psi (c,q) = \eu^{\iu S(c,q) / \hbar} ,
		\end{equation}
		and expand the complex phase~$ S $ in powers of the expansion parameter. We have
		\begin{equation}
			S = \sum_{n = 0}^{\infty} K^n S_n ,
		\end{equation}
		where for the $\hbar$ expansion
		\begin{equation}
		\label{eq:defKVil}
			K^n = \hbar^n
		\end{equation} 
		and for the $M$ expansion
		\begin{equation}
		\label{eq:defKKief}
			K^n = M^{1-n} .
		\end{equation}
		To obtain the factorized form of the wave function, we assume that each order of the expansion of $ S $ after the first can be separated as
		\begin{equation}
			S_n = \sigma_n(c) + \eta_n(c,q) ,\quad n \ge 1 .
		\end{equation}
		This way, we obtain
		\begin{equation}
			S = K^0 S_0 + P + Q, 
		\end{equation}
		where we have defined
		\begin{subequations}
		\begin{gather}
			P(c) = \sum_{n = 1}^{\infty} K^n \sigma_n ,\\
			Q(c,q) = \sum_{n = 1}^{\infty} K^n \eta_n .
		\end{gather}
		\end{subequations}
		We stress that the lowest order is slightly different in the two expansions; in fact, it can be seen that in the $\hbar$ expansion the total action is $ S = S_0 + P + Q$, while in the $M$ expansion it takes the form $S = M S_0 + P + Q$.
		The wave function then takes the BO-like form
		\begin{equation}
		\label{eq:PsiDecomposition}
			\Psi(c,q) = \psi(c) \chi(c,q)
		\end{equation}
		where for the $ \hbar $ expansion
		\begin{subequations}
		\begin{gather}
		\label{eq:expansionBackground}
			\psi(c) = \eu^{\iu (S_0 + P) / \hbar}\\
			\chi(c,q) = \eu^{\iu Q / \hbar} ,
		\end{gather}
		\end{subequations}
		and for the $ M $ expansion
		\begin{subequations}
		\begin{gather}
		\label{eq:expansionBackgroundKiefer}
			\psi(c) = \eu^{\iu (M S_0 + P) / \hbar}\\
			\chi(c,q) = \eu^{\iu M Q / \hbar} .
		\end{gather}
		\end{subequations}
		
		The background wave function is assumed to satisfy the WKB equation for the classical part alone, i.e.\ for the $ \hbar $ case
		\begin{equation}
		\label{eq:VilBackground}
			\left( -\hbar^2 \nabla_c^2 + U_c \right) \psi(c) = 0
		\end{equation}
		or for the $ M $ one
		\begin{equation}
		\label{eq:background_M}
			\left( -\frac{\hbar^2}{2 M} \nabla_g^2 + M V \right) \psi(g) = 0 .
		\end{equation}
		Here, the difference on the background between the expansions is evident: apart from numerical factors, the $ M $ expansion automatically forces the gravitational part to be exactly the classical one (limit of a vacuum universe).
		
		By substituting to $ \psi $ its expansion~\eqref{eq:expansionBackground}, this equation yields, order by order, the Hamilton-Jacobi equation for the classical action~$ S_0 $ (which must be a real function in order to give the correct classical limit, as discussed in~\cite{bib:kiefer-1991}) and the equations of the WKB expansion for each~$ \sigma_n $. We report the first orders here: for the $ \hbar $ expansion we have
		\begin{subequations}
		\begin{gather}
		\label{eq:HJ_h}
			(\nabla_c S_0)^2 + U_c = 0,\\
		\label{eq:WKB_h}
			2 \nabla_c S_0 \cdot \nabla_c \sigma_1 - \iu \nabla_c^2 S_0 = 0,\\
			2 \nabla_c S_0 \cdot \nabla_c \sigma_2 + (\nabla_c^2 \sigma_1)^2 - \iu \nabla_c^2 \sigma_1 = 0,\\
			\begin{split}
				2 \nabla_c S_0 \cdot \nabla_c \sigma_3 &+ 2 \nabla_c \sigma_1 \cdot \nabla_c \sigma_2\\
				&- \iu \nabla_c^2 \sigma_2 = 0 ,
			\end{split}
		\end{gather}
		\end{subequations}
		while for the $ M $ expansion
		\begin{subequations}
		\begin{gather}
		\label{eq:HJ_M}
			\frac{1}{2} (\nabla_g S_0)^2 + V = 0,\\
		\label{eq:WKB_M}
			\nabla_g S_0 \cdot \nabla_g \sigma_1 - \frac{\iu \hbar}{2} \nabla_c^2 S_0 = 0,\\
			\nabla_g S_0 \cdot \nabla_g \sigma_2 + \frac{1}{2} (\nabla_g^2 \sigma_1)^2 - \frac{\iu \hbar}{2} \nabla_g^2 \sigma_1 = 0,\\
			\begin{split}
				\nabla_g S_0 \cdot \nabla_g \sigma_3 &+ \nabla_g \sigma_1 \cdot \nabla_g \sigma_2\\
				&- \frac{\iu \hbar}{2} \nabla_g^2 \sigma_2 = 0,
			\end{split}
		\end{gather}
		\end{subequations}
		where in the scalar products the metric tensor is implied by the dot.
		
		The equation for the quantum subsystem is obtained by plugging Eq.~\eqref{eq:PsiDecomposition} into the WDW equation, using Eq.~\eqref{eq:WKB_h} or~\eqref{eq:WKB_M} to remove the classical part and dividing by~$ \psi $. We get, respectively,
		\begin{equation}
		\label{eq:quantum_h}
			2 \hbar^2 \nabla_c \ln \psi \cdot \nabla_c \chi = H_q \chi - \hbar^2 \nabla_c^2 \chi
		\end{equation}
		for the $\hbar$ expansion, or
		\begin{equation}
		\label{eq:quantum_M}
			2 \frac{\hbar^2}{M} \nabla_g \ln \psi \cdot \nabla_g \chi = H_q \chi - \frac{\hbar^2}{2 M} \nabla_g^2 \chi
		\end{equation}
		for the $M$ expansion.
		
		After substituting $ \psi $ with Eq.~\eqref{eq:expansionBackground} or~\eqref{eq:expansionBackgroundKiefer} respectively, and defining time through the dependence on the classical part as in~\cite{bib:vilenkin-1989,bib:kiefer-1991}, i.e.
		\begin{equation}
			\frac{\partial}{\partial \tau} = 2 \nabla_c S_0 \cdot \nabla_c
		\end{equation}
		in the $ \hbar $ expansion and
		\begin{equation}
		\label{eq:time_M}
			\frac{\partial}{\partial \tau} = \nabla_g S_0 \cdot \nabla_g
		\end{equation}
		in the $ M $ one, eqs.~\eqref{eq:quantum_h} and~\eqref{eq:quantum_M} yield the corrected Schr\"odinger equation
		\begin{equation}
		\label{eq:schrodingerCorrected}
			\iu \hbar \frac{\partial \chi}{\partial \tau} = H_q \chi - k_1 \iu \hbar \nabla_c P \cdot \nabla_c \chi - k_2 \nabla_c^2 \chi.
		\end{equation}
		Here, in the $ \hbar $ expansion $ k_1 = 2 $ and $ k_2 = \hbar^2 $, while in the $ M $ expansion $ k_1 = 1 $ and $ k_2 = - \hbar^2 / (2 M) $. The different numerical factors in the last equation and in the definition of time are due to to the additional factor~$ 2 $ together with~$ M $ in the $ M $~expansion, and not to any physical reason.
		
		It is important to notice that at orders~$ \ord{\hbar} $ and~$ \ord{M^0} $, Eq.~\eqref{eq:schrodingerCorrected} reduces to the exact Schr\"odinger equation for the quantum wave functional~$ \chi_1 $:
		\begin{equation}
			\iu \hbar \frac{\partial \chi_1}{\partial \tau} = H_q \chi_1 .
		\end{equation} 
		At orders~$ \ord{\hbar^2} $ and~$ \ord{1/M} $ the corrections to the standard dynamics will emerge: it is easy to find
		\begin{equation}
		\label{eq:schrodingerCorrected_Ordh^2}
			\iu \hbar \frac{\partial \chi_2}{\partial \tau} = H_q \chi_2 - \left( \iu k_1 \nabla_c \sigma_1 \cdot \nabla_c + k_2 \nabla_c^2 \right) \chi_2 ,
		\end{equation}
		where in the $ \hbar $ expansion $ k_1 = 2 \hbar^2 $ and $ k_2 = \hbar^2 $, while in the $ M $ expansion $ k_1 = \hbar / M $ and $ k_2 = \hbar^2 / (2 M) $. The corrective terms are of the same kind of those in~\cite{bib:kiefer-1991} i.e.\ they are not unitary. This result shows that, if we restrict the classical subspace to the geometrical variables only, the $ \hbar $ expansion yields precisely the same results of the $ M $ expansion, also at the quantum gravity order.
		
		Following a procedure described in~\cite{bib:kiefer-2018}, Eq.~\eqref{eq:schrodingerCorrected} can be written in a nicer form. We assume the existence of a total (in general not Hermitian) Hamiltonian operator~$ \mathrm{H} $ such that
		\begin{equation}
			\iu \hbar \frac{\partial \chi}{\partial \tau} = \mathrm{H} \chi ,
		\end{equation}
		and we also assume that
		\begin{equation}
			\nabla_c \chi = \alpha (c) \nabla_c S_0 ,
		\end{equation}
		which is some sort of adiabatic approximation. The HJ equations~\eqref{eq:HJ_h} and~\eqref{eq:HJ_M} give
		\begin{equation}
			\alpha = - \frac{1}{2 k U_c} \frac{\partial \chi}{\partial \tau} = \frac{\iu}{2 \hbar k U_c} \mathrm{H} \chi ,
		\end{equation}
		where in the $ \hbar $~expansion $ k = 1 $, while in the $ M $~expansion~$ U_c = M V $ and $ k = 1 / M $, such that~$ k U_c = V $. Using eqs.~\eqref{eq:WKB_h} and~\eqref{eq:WKB_M}, the corrected Schr\"odinger equation~\eqref{eq:schrodingerCorrected} becomes
		\begin{subequations}
		\begin{gather}
			\begin{split}
			\label{eq:schrodingerCorrected_v2}
				\iu \hbar \frac{\partial \chi}{\partial \tau} &\equiv \mathrm{H} \chi = H_q \chi\\
				&\hphantom{=} - \frac{1}{4 k_1 U_c} \left( \mathrm{H}^2 + \iu \hbar \frac{\partial \mathrm{H} }{\partial \tau} - \iu \hbar K \mathrm{H} \right) \chi
			\end{split}\\
			K = \frac{1}{U_c} \frac{\partial U_c}{\partial \tau} - \frac{\iu k_2}{\hbar} \sum_{n=2}^\infty k_3^n \frac{\partial \sigma_n}{\partial \tau} ,
		\end{gather}
		\end{subequations}
		where in the $ \hbar $~expansion $ k_1 = k_2 = 1 $ and $ k_3 = \hbar $, while in the $ M $~expansion $ U_c = M V $, $ k_1 = 2 $, $ k_2 = 2 M $ and $ k_3 = 1 / M $. We remark that $ \mathrm{H} $~is an abstract Hamiltonian operator containing $ H_q $ and all the corrections at every order. These expressions show even more the equivalence of the two expansions, except for the initial assumptions discussed above.
		
		The procedure we just performed is the generalization of that used in~\cite{bib:kiefer-1991} to derive its Eq.~(42) and based on the decomposition of contributions tangential and orthogonal to the hypersurfaces $ S_0 = \const $. The use of eqs.~\eqref{eq:WKB_h} and~\eqref{eq:WKB_M} causes the sum in the expression of $ K $ used here to begin from~$ n = 2 $. At the quantum gravity order~$ \ord{\hbar^2} $ and $ \ord{1 / M} $, Eq.~\eqref{eq:schrodingerCorrected_v2} yields the final equation of~\cite{bib:kiefer-1991}, that is its Eq.~(42). At higher orders, the quantum gravity corrections not only arise from the $ \nabla_c^2 $ term in Eq.~\eqref{eq:schrodingerCorrected}, but also from the term containing~$ P $. As noted in~\cite{bib:kiefer-2018}, the same result can be obtained by considering $ \sigma_n $, the classical potential $ V $ (or $ U_c $, in the $ \hbar $ expansion) and $ \chi $ depending only on $ \tau $ from the beginning and dropping all the components of the supermetric corresponding to the geometric subspace with the exception of the $ \mathcal{G}_{\tau\tau} $ component.
		
		Let us now assess the situation. Both the $ \hbar $ and the $ M $ expansions recover the already established theories through a HJ equation for GR, that fixes a background, and a Schr\"odinger equation in curved space-time for quantum mechanics. The $ \hbar $ expansion is more general, since it admits backgrounds generated by matter sources and quantum geometry. At the quantum gravity order, both expansions yield non-Hermitian corrections that break the unitarity of the theory. A further common feature of the two approaches is that the backreaction of the quantum subsystem on the background is not present. The inclusion of such a nonadiabatic effect would allow for quantum gravitational effects on the semiclassical sector.
		
	\subsection{Nonunitarity in the revisited Planck mass expansion\label{ssec:non-unit_Plmassexp}}
		The problem of nonunitarity is a huge drawback of the model. Nonetheless, in~\cite{bib:kiefer-2016}, the resulting nonunitary dynamics of the quantum matter components is applied to compute the quantum-gravitational corrections to the power spectra of gauge-invariant scalar and tensor perturbations during the inflationary phase of the Universe. In~\cite{bib:kiefer-2018}, working in the $M$ expansion, the authors develop a procedure to make the quantum gravity Hamiltonian a Hermitian operator. We here show that such procedure is based on wrong assumptions.
		
		Let us briefly apply such procedure to the simple case of one geometric variable, that we identify with the time~$ \tau $ from the beginning. Once we use the ansatz $ \Psi(\tau,m) = \psi(\tau) \chi(\tau,m) $, the WDW equation~\eqref{eq:WDW} reads
		\begin{equation}
			\frac{\hbar^2}{M} \mathcal{G}_{\tau\tau} \partial_\tau \ln \psi \, \partial_\tau \chi = H_m \chi - \frac{\hbar^2}{2 M} \mathcal{G}_{\tau\tau} \partial_\tau^2 \chi +\rho_\psi \chi ,
		\end{equation}
		where the background term
		\begin{equation}
			\rho_\psi = \frac{1}{\psi} \left[ - \frac{\hbar^2}{2 M} \mathcal{G}_{\tau\tau} \partial_\tau^2 + M V \right] \psi
		\end{equation}
		corresponds to the quantity set to zero in Eq.~\eqref{eq:background_M}.
		
		Differently from~\cite{bib:kiefer-1991}, in~\cite{bib:kiefer-2018}, after writing $ \psi_0 = \exp{(\iu M S_0 / \hbar)} $, the background term $ \rho_{\psi_0} $ is required to be of order~$ \ord{M^0} $. In order to satisfy this request, the HJ equation
		\begin{equation}
			\frac{1}{2} \mathcal{G}_{\tau\tau} (\partial_\tau S_0)^2 + V = 0
		\end{equation}
		has to hold at order~$ \ord{M} $. Hence, the expression of $ \rho_{\psi_0} $ at order~$ \ord{M^0} $ is
		\begin{equation}\label{eq:Kief18-back0}
			\rho_{\psi_0} = - \iu \hbar \frac{\partial_\tau V}{2 V} .
		\end{equation}
		Using Eq.~\eqref{eq:Kief18-back0} and assuming the existence of an abstract Hamiltonian operator~$ \mathrm{H} $ similar to that defined in~\eqref{eq:schrodingerCorrected_v2}, we find
		\begin{equation}
		\label{eq:Kief18-WDW}
		\begin{split}
			\iu \hbar \partial_\tau \chi_0 &\equiv \mathrm{H} \chi_0 = H_m \chi_0\\
			&\hphantom{=} -\frac{\iu \hbar}{2} \frac{\partial_\tau V}{V} \chi_0 -\frac{1}{4 M V} \left[ \mathrm{H}^2 + \iu \hbar \partial_\tau \mathrm{H} \right] \chi_0 ,
		\end{split}
		\end{equation}
		where $ \chi_0 $ is the quantum wave function, such that $ \Psi = \psi_0 \chi_0 $. This equation still exhibits non-Hermitian corrections. To deal with them, in~\cite{bib:kiefer-2018}, the authors assume the existence of two eigenvalue functions, $ E(\tau) $ (complex) and $ \epsilon(\tau) $ (real) such that
		\begin{subequations}
		\label{eq:Kief18-Autov}
		\begin{gather}
			\mathrm{H} \chi_0 = E(\tau) \chi_0\\
			H_m \chi_0 = \epsilon(\tau) \chi_0 ,
		\end{gather}
		\end{subequations}
		and expand them in powers of~$ 1 / M $. Written in terms of these expansions, the WDW equation~\eqref{eq:Kief18-WDW} yields, at each order, an expression for the eigenvalue of the abstract Hamiltonian operator. Let us report the first two orders ($ M^0 $ and $ 1 / M $):
		\begin{subequations}
		\label{eq:Kief18-Autov2}
		\begin{gather}
			E^{(0)} = \epsilon^{(0)} - \frac{\iu \hbar}{2} \frac{\partial_\tau V}{V}\\
			\begin{split}
				E^{(1)} &= \epsilon^{(1)} - \frac{1}{4 V} \left[ \left( \epsilon^{(0)} \right)^2 + \hbar^2 \frac{\partial_\tau^2 V}{2 V} \vphantom{\left( \frac{\partial_\tau V}{V} \right)^2} \right.\\
				&\hphantom{=}\left. - \frac{3 \hbar^2}{4} \left( \frac{\partial_\tau V}{V} \right)^2 \right] - \frac{\iu \hbar}{4} \partial_\tau \left( \frac{\epsilon^{(0)}}{V} \right) .
			\end{split}
		\end{gather}
		\end{subequations}
		Defining
		\begin{equation}
		\label{eq:redef1}
			\chi_1 = \eu^{-\frac{1}{\hbar} \int \imaginaryPart ( E^{(0)}) \di \tau} \chi_0 = \eu^{\int \frac{\partial_\tau V}{2 V}} \chi_0 ,
		\end{equation}
		and substituting into Eq.~\eqref{eq:Kief18-WDW} we find
		\begin{equation}
			\iu \hbar \partial_\tau \chi_1 = H_m \chi_1 .
		\end{equation} 
		The time derivative of the redefined quantum state contributes with a term that exactly compensates the non-Hermitian correction on the right-hand side of Eq.~\eqref{eq:Kief18-WDW}, due to Eq.~\eqref{eq:Kief18-back0} at this order~$ \ord{M^0} $. The background term must now be calculated for a $ \psi_1 $ defined in such a way that $ \Psi = \psi_1 \chi_1 $, i.e.\
		\begin{equation}
			\psi_1 = \eu^{-\int \frac{\partial_\tau V}{2 V}} \psi_0 = \eu^{\iu M S_0 / \hbar + \sigma_1} ,
		\end{equation}
		where $ \sigma_1 = - \ln V / 2 $. By doing so, we find that $ \rho_{\psi_1} $ vanishes at order $ \ord{M^0} $, yielding the continuity equation
		\begin{equation}
			\partial_\tau^2 S_0 + \partial_\tau S_0 \partial_\tau \sigma_1 = 0 .
		\end{equation}
		We can easily see that this equation vanishes naturally. Thus, $ \rho_{\psi_1} $ is of order $ \ord{1/M} $ and is given by the expression
		\begin{equation}
		\label{eq:Kief18-back1}
			\rho_{\psi_1} = \frac{\hbar^2}{4 M V} \left[ \frac{3}{4} \left( \frac{\partial_\tau V}{V} \right)^2 - \frac{\partial_\tau^2 V}{2 V} \right] .
		\end{equation}
		
		The same steps can be followed at order $ \ord{1/M} $, including the term in Eq.~\eqref{eq:Kief18-back1} into Eq.~\eqref{eq:Kief18-WDW} and redefining the quantum state as
		\begin{equation}
		\label{eq:redef2}
			\chi_2 = \eu^{-\frac{1}{M \hbar} \int \imaginaryPart \left( E^{(1)} \right) \di \tau} \chi_1 .
		\end{equation}
		The corrected Schr\"odinger equation will have only the Hermitian part of the Hamiltonian operator~$ \mathrm{H} $, exhibiting unitary evolution. The background term calculated for a $ \psi_2 $ such that $ \Psi = \psi_2 \chi_2 $ will not vanish naturally at this order, as an effect of the backreaction of the quantum subsystem.
		
		This procedure is based on the nice idea that the non-Hermitian part of the operator $ \mathrm{H} $ may be eliminated from the dynamical equation of the quantum subsystem by suitable redefinitions of the wave functions in the product $ \Psi = \psi \chi $. However, the $ \mathrm{H} $ operator is unknown in general and can only be constructed order by order; moreover, in order to redefine the wave functions through phase factors, one has to use the eigenvalues of $ \mathrm{H} $. The problem lies in eqs.~\eqref{eq:Kief18-Autov}, which implies that the operators $ H_m $ and $\mathrm{ H} $ commute at every order and can be diagonalized simultaneously. Unfortunately, this is clearly not true at every order, as one can see from the expression of $ E^{(1)} $ in eqs.~\eqref{eq:Kief18-Autov2}. Indeed, $ E^{(1)} $ contains $ \epsilon^{(0)} $ and its time derivative $ \partial_\tau \epsilon^{(0)} $, meaning that the Hamiltonian $\mathrm{H} $ at the order $ \ord{1/M} $ contains the matter Hamiltonian $ H_m $ and its time derivative $ \dot{H}_m $, coherently with Eq.~\eqref{eq:schrodingerCorrected_v2} expressed at order~$ \ord{1/M} $. In the general case, it is not true that $ H_m $ and $ \dot{H}_m $ commute: the reason is that, in principle, the expression of~$ \dot{H}_m $ contains coordinate and conjugate momenta operators not commuting with $H_m$.
		
		To convince ourselves about this, let us consider a Friedmann-Robertson-Walker (FRW) model with cosmological constant and a scalar field as matter component. The Hamiltonian constraint reads
		\begin{subequations}
		\label{eq:toy}
		\begin{gather}
			H_\mathrm{FRW} = - \frac{G}{32 c^3 \pi a} p_a^2 + \frac{c}{4 \pi^2 a^3} p_\phi^2 - V\\
			V(a;\Lambda) = \frac{3 \pi c^3}{4 G} \left( a - \frac{\Lambda}{3} a^3 \right) ,
		\end{gather}
		\end{subequations}
		where $ V $ is the FRW superpotential. An important remark is that the conjugated momenta to the volume of the universe $ a $ is proportional to the time derivative of~$ a $:
		\begin{equation}
			p_a \sim \frac{a}{N} \frac{\di a}{\di t} = a \partial_\tau a .
		\end{equation}
		The matter Hamiltonian of this simple model is just
		\begin{equation}
			H_m = \frac{c}{4 \pi^2} a^{-3} p_\phi^2
		\end{equation}
		and its time derivative yields
		\begin{equation}
			\partial_\tau H_m = - \frac{3 c}{4 \pi^2} a^{-4} \partial_\tau a p_\phi^2 \sim a^{-5} p_a p_\phi .
		\end{equation}
		The appearance of $ p_a $ in $ \partial_\tau H_m $ clearly leads to $ [H_m, \partial_\tau H_m] \neq 0 $. 
		
		A further issue of the procedure followed in~\cite{bib:kiefer-2018} concerns the absence of gauge invariance in this approach: even if the total wave function $ \Psi $ is invariant under the redefinitions performed on $ \psi $ and $ \chi $, the equations of motion are not, differently from what happens in~\cite{bib:bertoni-1996}. Thus, such redefinitions cannot be fully justified on theoretical grounds.
		
\section{Exact decomposition with extended BO approach\label{sec:BOExpansion}}
	Another attempt to treat a quantum subsystem on a WKB background is provided by~\cite{bib:bertoni-1996,bib:venturi-2017}. In these works, the authors develop a decomposition in classical and quantum variables through an extended BO approach, that is more accurate than the traditional one and is largely used in chemistry~\cite{bib:gidopoulos-2005,bib:abedi-2010,bib:scherrer-2017,bib:alonso-2013}, where it finds experimental verification. 
	
	\subsection{BO decomposition and WKB expansion of the matter-gravity problem\label{ssec:BO-Venturi}}
		The approach of~\cite{bib:bertoni-1996}, set in the minisuperspace, is based on the exact decomposition of the wave function
		\begin{subequations}
		\label{eq:bertoni-decomposition}
		\begin{gather}
			\Psi(c, q) = \psi(c) \chi(q; c)\\
			\langle \chi | \chi \rangle = \int \chi^{*}(q; c) \chi(q; c) \di q = 1 ,
		\end{gather}
		\end{subequations}
		from which the equations for the background and for the quantum subsystem are obtained, respectively, by averaging the WDW equation~\eqref{eq:WDW} over the quantum functional~$ \chi $ and by subtracting the resulting equation to the initial WDW equation:
		\begin{subequations}
		\label{eq:bertoni-eq1set}
		\begin{gather}
			\begin{split}
				&\left[ - \frac{\hbar^2}{2 M} \left( \D^2 + \avg{\Db^2} \right) \right.\\
				&\qquad \left. \vphantom{\frac{\hbar^2}{2 M}} + M V + \avg{H_q} \right] \psi = 0
			\end{split}\\
			\begin{split}
				&\left[ - \frac{\hbar^2}{2 M} \left( \Db^2 - \avg{\Db^2} + 2 \D \ln \psi \Db \right) \right.\\
				&\qquad+ \left. \vphantom{\frac{\hbar^2}{2 M}} H_q - \avg{H_q} \right] \chi = 0 .
			\end{split}
		\end{gather}
		\end{subequations}
		Here we have used the definitions:
		\begin{subequations}
		\label{eq:bertoni-derivatives}
		\begin{gather}
			\avg{\mathcal{O}} = \langle \chi | \mathcal{O} | \chi \rangle\\
		\label{eq:bertoni-connect}
			A = - \iu \hbar \avg{\nabla_c}\\
			- \iu \hbar \D = - \iu \hbar \nabla_c + A\\
			- \iu \hbar \Db = - \iu \hbar \nabla_c - A ,
		\end{gather}
		\end{subequations} 
		where the quantity $ A $ plays the role of a Berry connection and $ \D, \Db $ are covariant derivatives.
		
		This approach has some nice properties, for which it may be preferred to the one analyzed in the previous section. First of all we note that, given the ansatz~\eqref{eq:bertoni-decomposition}, if the total wave function is normalized to unit then, as a bonus, it follows straightforwardly that the background wave function is normalized as well. Moreover eqs.~\eqref{eq:bertoni-eq1set} can be more generally derived from a variational principle~\cite{bib:gidopoulos-2005,bib:abedi-2010}. Another property of eqs.~\eqref{eq:bertoni-decomposition} is that they imply no freedom to the decomposition of the total wave function into classical and quantum components, except for a phase factor depending on the classical variables only. Equation~\eqref{eq:bertoni-eq1set} is invariant under such a phase change, due to the covariant derivatives. Hence the decomposition~\eqref{eq:bertoni-decomposition} is gauge invariant. The covariant derivatives~$ \D,\Db $ can be absorbed in the wave functions through the redefinitions
		\begin{subequations}
		\label{eq:bertoni-tildeVars}
		\begin{gather}
			\psi = \eu^{- \frac{\iu}{\hbar} \int A \di c} \widetilde{\psi}\\
			\chi = \eu^{\frac{\iu}{\hbar} \int A \di c} \widetilde{\chi} ,
		\end{gather}
		\end{subequations}
		leading to a second set of equations
		\begin{subequations}
		\label{eq:bertoni-eq2set}
		\begin{gather}
			\begin{split}
				&\left[ - \frac{\hbar^2}{2 M} \left( \nabla_c^2 + \avgt{\nabla_c^2} \right) \right.\\
				&\qquad\left. \vphantom{\frac{\hbar^2}{2 M}} + M V + \avg{H_q} \right] \widetilde{\psi} = 0
			\end{split}\\
		\label{eq:bertoni_tildeQuantum}
			\begin{split}
				&\left[ - \frac{\hbar^2}{2 M} \left( \nabla_c^2 - \avgt{\nabla_c^2} + 2 \nabla_c \ln \widetilde{\psi} \cdot \nabla_c \right) \right.\\
				&\qquad + \left. \vphantom{\frac{\hbar^2}{2 M}} H_q - \avg{H_q} \right] \widetilde{\chi} = 0 ,
			\end{split}
		\end{gather}
		\end{subequations}
		where we defined the average over the wave function $ \widetilde{\chi} $ as $ \avgt{\mathcal{O}} = \langle \widetilde{\chi} | \mathcal{O} | \widetilde{\chi} \rangle $ and we used the property $ \avgt{H_q} = \avg{H_q} $, since $ H_q $ acts only on the quantum variables.
		
		Here our analysis will depart from that carried out in~\cite{bib:bertoni-1996}, that is affected by a few evident issues. The authors proceed performing the WKB expansion on the background function $ \widetilde{\psi} $ to define the semiclassical time as before~\eqref{eq:time_M}, obtaining the presence of backreaction in the HJ equation; simultaneously, they perform the following rescaling of the quantum wave function:
		\begin{equation}
		\label{eq:chi_s}
			\widetilde{\chi} = \eu^{\frac{\iu}{\hbar} \int \avg{H_m} \di \tau} \chi_s ,
		\end{equation}
		in order to find the corrected Schr\"odinger equation
		\begin{equation}
		\label{eq:bertoni-schrodinger}
			\begin{split}
				&( H_q - \iu \hbar \partial_\tau ) \chi_s = \eu^{- \frac{\iu}{\hbar} \int \avg{H_q} \di \tau - \frac{\iu}{\hbar} \int A \di c}\\
				&\quad \times \frac{\hbar^2}{2 M} \left[ \Db^2 - \avg{\Db^2} + 2 ( \D \ln N ) \Db \right] \chi .
			\end{split}
		\end{equation}
		
		For further discussion on the phase transformations here performed, see Appendix~\ref{APP-phase}.
		
		Finally, they show the unitarity of the theory through
		\begin{equation}
		\label{eq:bertoni-unitarity}
			\iu \hbar \partial_\tau \langle \chi_s | \chi_s \rangle = 0 .
		\end{equation}
		We now have to remark some critical points in the approach just described. One evident problem with this procedure is that the quantum wave function is given by $ \chi_s $ and the semiclassical wave function by $ \widetilde{\psi} $, while their product should yield the total wave function: this implies a breaking of the gauge symmetry of the theory, that was one of the merits of this formalism.
		
		However, the most serious issue is that the Schr\"odinger equation~\eqref{eq:bertoni-schrodinger} contains derivatives with respect to the background variables, which in turn contain the WKB time: these derivatives must be clearly expressed and analysed as suggested by~\cite{bib:kiefer-1991} and discussed in Sec.~\ref{sec:WKB}.
		
		A further issue is that, even if it was correct, the procedure followed in~\cite{bib:bertoni-1996} would not really show the unitarity of the theory: in fact, Eq.~\eqref{eq:bertoni-unitarity} vanishes only if one takes the norm of the states, but the cancellation fails for different quantum states. This means that a proper dynamical Hilbert space cannot be built in this approach, since a conserved scalar product cannot be defined for all the states.
		
	\subsection{Enhancement and nonunitarity of the BO decomposition\label{ssec:EnhancementBO}}
		We will now improve this method in order to deal with the open issues just discussed, and we will show that, even when this formalism is used properly, nonunitarity still affects the dynamics of the quantum subsystem at the quantum gravity order.
		
		First, we define the background wave functional~$ \psi_s $ associated with~$ \chi_s $ through
		\begin{equation}
		\label{eq:psi_s}
			\widetilde{\psi} = \eu^{- \frac{\iu}{\hbar} \int \avg{H_q} \di \tau} \psi_s ,
		\end{equation}
		in such a way that the total wave function reads
		\begin{equation}
			\Psi = \psi \chi = \widetilde{\psi} \widetilde{\chi} = \psi_s \chi_s .
		\end{equation}
		In order to ease the comparison with~\cite{bib:kiefer-2018} and the original results of~\cite{bib:bertoni-1996}, which is implemented in the minisuperspace, we use the same formalism and we perform the semiclassical expansion of the background wave functional in~$ M $
		\begin{equation}
			\psi_s = \eu^{\iu M (S_0 + P) / \hbar} ,
		\end{equation}
		where
		\begin{equation}
			P = \sum_{n=1}^\infty M^{-n} \sigma_n .
		\end{equation}
		We now decompose the quantity $ P $ in its real and imaginary parts as $ P = \zeta - \iu \rho $, such that
		\begin{equation}
			\psi_s = \eu^{M \rho / \hbar} \eu^{\iu M (S_0 + \zeta) / \hbar} ,
		\end{equation}
		where
		\begin{subequations} 
		\begin{gather}
			\realPart (P) \equiv \zeta = \frac{1}{M} \zeta_1 + \frac{1}{M^2} \zeta_2 + \ldots\\
			-\imaginaryPart (P) \equiv \rho = \frac{1}{M} \rho_1 + \frac{1}{M^2} \rho_2 + \ldots 
		\end{gather}
		\end{subequations}
		We define the WKB time as usual as
		\begin{equation}
		\label{eq:time}
			\iu \hbar \partial_\tau = \iu \hbar \nabla_c S_0 \cdot \nabla_c
		\end{equation}
		and Eq.~\eqref{eq:bertoni-schrodinger} at order~$ \ord{M^0} $ yields the Schr\"odinger equation
		\begin{equation}
		\label{eq:schrodinger}
			(- \iu \hbar \partial_\tau + H_q) \chi_s = 0 ,
		\end{equation}
		as expected.
		
		The first interesting difference from~\cite{bib:bertoni-1996} is that the backreaction cancels out from the equations of the background expansion: we find at order~$ \ord{M} $ the usual HJ equation 
		\begin{equation}
		\label{eq:ordM-classical}
			\frac{1}{2} (\nabla_c S_0)^2 + V = 0 ,
		\end{equation}
		and at order~$ \ord{M^0} $ 
		\begin{equation}
		\label{eq:ordM0-classical}
			\begin{split}
				&- \frac{\iu \hbar}{2} \nabla_c^2 S_0 + \nabla_c S_0 \cdot \nabla_c \zeta_1 - \iu \nabla_c S_0 \cdot \nabla_c \rho_1\\
				&\quad - \nabla_c S_0 \cdot \nabla_c \int \avg{H_q} \di \tau + \avg{H_q} = 0 ,
			\end{split}
		\end{equation}
		where the last two terms cancel because of the definition of time. Thus, the backreaction has shifted from the HJ equation to the continuity equation, where it is canceled out by our redefinition of the background functional~\eqref{eq:psi_s}. The same would have happened in the $ \hbar $ expansion, because of the hypothesis of smallness of the quantum subsystem.
		
		After separating the real and imaginary parts, Eq.~\eqref{eq:ordM0-classical} yields
		\begin{subequations}
		\label{eq:background-ord1}
		\begin{gather}
		\label{eq:background-ord1-eta}
			\frac{\hbar}{2} \nabla_c^2 S_0 + \nabla_c S_0 \cdot \nabla_c \rho_1 = 0\\
		\label{eq:S1}
			\nabla_c S_0 \cdot \nabla_c \zeta_1 = 0 .
		\end{gather}
		\end{subequations}
	
		The first equation corresponds exactly to Eq.~\eqref{eq:WKB_M}, while the second points out that $ \zeta_1 $ has no dynamical relevance: through Eq.~\eqref{eq:time}, Eq.~\eqref{eq:S1} reads
		\begin{equation}
			\partial_\tau \zeta_1 = 0 .
		\end{equation}
		
		Until now, we recovered precisely the results of~\cite{bib:kiefer-1991} (and equivalently~\cite{bib:vilenkin-1989}), but with the adoption of the more advanced formalism of~\cite{bib:bertoni-1996}.
		
		To investigate the quantum gravity order~$ \ord{1/M} $, we restrict our analysis, for simplicity, to the case of a cosmological model with a single gravitational degree of freedom, which will be denoted as~$ \alpha $. This will keep us from dealing with the projection of the gradients in the geometrical indices with respect to the $ S_0 = \const $ hypersurfaces. The procedure is valid only if~$ \spR \neq 0 $; otherwise, we would have $V=0$ and this would give trouble in the next steps. The detailed calculations at this order are summarized in Appendix~\ref{APP-BODecomposition}.

		The corrected Schr\"odinger equation up to order~$ \ord{1/M} $ is
		\begin{equation}
		\label{eq:finalSchrBO}
			\begin{split}
				&\iu \hbar \partial_\tau \chi_s = H_q \chi_s - \frac{1}{4 M V} \left[ \left( H_q^2 - \avg*{H_q^2} \right) \vphantom{\frac{\dot{V}}{V}} \right.\\
				&\quad+ \left. \iu \hbar (\dot{H}_q - \avg{\dot{H}_q}) - \iu \hbar \frac{\dot{V}}{V} (H_q - \avg{H_q}) \right] \chi_s .
			\end{split}
		\end{equation}
		The last equation is the equivalent of Eq.~\eqref{eq:schrodingerCorrected_v2} up to order~$ \ord{1/M} $, i.e.\ of Eq.~(42) of~\cite{bib:kiefer-1991}, but in the framework of~\cite{bib:bertoni-1996}.

		The computation of the corresponding equation for the background wave function $\psi$ is reported in Appendix~\ref{APP-BODecomposition}.
		
		We note in \eqref{eq:finalSchrBO} that the non-Hermiticity of the quantum gravity Hamiltonian is still a problem, unless one takes the norm of a state, hence Eq.~\eqref{eq:bertoni-unitarity}. In this case, differently from~\cite{bib:kiefer-1991}, all quantum gravity corrections vanish and this may be interpreted as a prediction of this approach.
		
\section{Unitarity with use of the kinematical action\label{sec:AzioneCin}}
	In this section we develop a proposal to solve the nonunitarity problem, based on a WKB expansion in the $M$ parameter as in~\cite{bib:kiefer-1991} but using a different construction of time, meaning a different physical clock for the gravity-matter system in the considered WKB separation of the dynamics. The previous works here discussed are all based on the definition of a classical time, i.e.\ constructed with the dependence of the subsystem on the classical variables. This choice seems to be the origin point of such nonunitary corrections, as discussed in the next subsection.
	
	For this reason, we here propose a model with introduction of the so-called kinematical action~\cite{bib:kuchar-1981}, see also \cite{bib:montani-2002}, which allows a covariant construction of parabolic constraints for quantum matter fields on a curved classical background. This term will be used to construct the time parameter of the theory, resulting in a quantum matter dynamics influenced by quantum-gravity corrections that present a unitary character.
	
	\subsection{Origin of the nonunitary corrections\label{ssec:OriginNonUn}}
		The critical analysis illustrated up to here is focused on the emerging problem of nonunitarity of the dynamics for quantum matter field on the WKB expanded gravitational background. This characteristic is present in the model~\cite{bib:vilenkin-1989}, that is truthfully expanded only up to order $\hbar$, and in the different proposal~\cite{bib:kiefer-1991} emerging at order $1/M$ in the expansion, as well as in the work~\cite{bib:bertoni-1996} based on the extended BO decomposition. All these approaches focus on the use of semiclassical variables as a clock for quantum matter: they construct the time derivative using the dependence of the matter wave function in terms of the classical generalized coordinates.
		
		However, it appears from the previous analyses that the most important term responsible for the nonunitarity of the models is the classical Laplacian~$ \nabla_c^2 $. Be it through some adiabatic assumption on the quantum wave function, some projection parallel and orthogonal to the hypersurfaces~$ S_0 = \const $, or simply by having time as the classical variable from the beginning, at some point that Laplacian generates~$ \nabla_\tau^2 \chi $. This is the crucial point that always generates nonunitarity, because it holds
		\begin{equation}
		\label{eq:origin_nonunit}
			- \hbar^2 \nabla_\tau^2 \chi = \iu \hbar \partial_\tau (H \chi) = \iu \hbar \dot{H} \chi + H^2 \chi ,
		\end{equation}
		where the incriminated term arises. Thus, until time is defined through~$ \nabla_c $, the model is probably doomed to find non-Hermitian corrections to the Hamiltonian at the quantum gravity level.
		
		For this reason, we propose here a different definition of time by using the kinematical action.
		
	\subsection{The kinematical action\label{ssec:IntrodAzioneCin}}
		The kinematical action was first introduced in~\cite{bib:kuchar-1981} as a tool to maintain the constraint equations of a quantum system by adding variables in the Lagrangian and Hamiltonian formalisms. We will now see this procedure applied to the case of scalar fields in a curved background.
		
		The kinematical action in the ADM representation reads:
		\begin{equation}
		\label{kinAction}
			S_{k} = \int \di^4 x (p_{\mu} \partial_t y^{\mu} - N^{\mu} p_{\mu}) ,
		\end{equation}
		where the coordinates $y^{\mu}$ are those defining the parametric equations of the hypersurfaces in the ADM splitting, as in $y^{\mu} = y^{\mu} (x^i; x^0)$, and $p^{\mu}$ are the associated momenta. The additional equations of motion, obtained by variations of $y^{\mu}$, $p_{\mu}$ and $N^{\mu}$, show that the momenta $p_{\mu}$ are trivial (equal to 0) and ensure that the physical meaning of the deformation vector $N^{\mu}$ is recovered:
		
		\begin{equation}
		\label{defVector}
			N^{\mu} = \partial_t y^{\mu} = N n^{\mu} + N^i b_i^{\mu} .
		\end{equation}
		
		This term also gives additional contributions to the total super-Hamiltonian and supermomentum constraints of the system:
		\begin{subequations}
		\label{Hkin}
		\begin{gather}
			\mathcal{H}^{k} = n^{\mu} p_{\mu}\\
			\mathcal{H}_i^{k} = b_i^{\mu} p_{\mu}
		\end{gather}
		\end{subequations}
		which are key elements to define a meaningful time variable for the matter field dynamics, in a different way than the works analyzed above.
		
		To show this, let us consider a massive scalar field immersed in a given gravitational background (assigned metric tensor). By use of the ADM variables the action can be written as:
		\begin{equation}
			S_{\phi} = \int \di x^0 \di^3 x \left( \pi \dot{\phi} -N \mathcal{H}^{\phi} - N^i \mathcal{H}_i^{\phi} \right) ,
		\end{equation}
		where $\pi$ is the momentum conjugated to the scalar field:
		\begin{equation}
			\pi = \left( -\frac{1}{N^2} \dot{\phi} +2 \frac{N^i}{N^2} \partial_i \phi \right) N\sqrt{h} .
		\end{equation}
		The super-Hamiltonian of the scalar field reads:
		\begin{equation}
		\label{Hphi}
			\mathcal{H}^{\phi} = \frac{1}{2\sqrt{h}} \pi^2 + \frac{1}{2} \sqrt{h} \nabla \phi\cdot \nabla \phi + \frac{1}{2}\sqrt{h} m^2 \phi^2 ,
		\end{equation}
		where in a short notation $\nabla \phi\cdot \nabla \phi \equiv h^{ij} \partial_i \phi \partial_j \phi$ , and the supermomentum of the scalar field takes the form:
		\begin{equation}
		\label{Hiphi}
			\mathcal{H}_i^{\phi} = (\partial_i \phi ) \pi .
		\end{equation}
	
		We notice that this way, the lapse function and shift vector $N$ and $N^i$ are assigned functions up to a restriction of the initial Cauchy problem: they are not to be varied, thus the physical definition of the ADM foliation on the background is lost.

		However, by adding the term \eqref{kinAction}, independent from the metric and matter field variables, we have:
		\begin{equation}
			\begin{split}
				&S^{\mathrm{tot}} = S^{\phi} + S^{k} = \int \di x^0 \di^3 x \left[ p_{\mu} \dot{y}^{\mu} + \pi \dot{\phi} \vphantom{\left(\mathcal{H}^{\phi}_i + \mathcal{H}^{k}_i \right)} \right.\\
				&\quad \left. - N \left(\mathcal{H}^{\phi} + \mathcal{H}^{k}\right) - N^i \left(\mathcal{H}^{\phi}_i + \mathcal{H}^{k}_i \right) \right] .
			\end{split}
		\end{equation}
		Thus the dynamics of the scalar field is left unchanged, but the definition of the deformation vector is recovered, as shown in~\eqref{defVector}, and the super-Hamiltonian and supermomentum constraints become:
		\begin{subequations}
		\label{eq:newConstrKinAction}
			\begin{gather}
				\mathcal{H}^{\phi} = - \mathcal{H}^{k} = -p_{\mu} n^{\mu}\\
				\mathcal{H}^{\phi}_i = - \mathcal{H}^{k}_i = - p_{\mu} b^{\mu}_i .
			\end{gather}
		\end{subequations}
		
		It is clear then that the addition of the kinematical action allows to recover the definition of the deformation vector and so the structure of the space-time foliation, which would otherwise be lost in this case. As a matter of fact, the kinematical action restores the geometrical meaning of the lapse function and of the shift vector, \emph{de facto} allowing their variation during the implementation of the variational principle.
		
		It follows that the quantum dynamics of the field is characterized by parabolic constraints, linear in the momentum canonically conjugate to the four-dimensional variables, thought as fields depending on the slicing space-time variables. In the canonical quantization procedure, the momenta $p^{\mu}$ will be transformed into derivative operators, and they will be crucial in the construction of the time derivative. We will now show that this procedure can be applied to the case of interest to obtain a quantum matter field dynamics without nonunitary terms arising from the previous proposals, which would prevent the predictability of the theory.
		
	\subsection{Unitary evolution with use of the kinematical action\label{ssec:UnitaryEv}}
		We now construct the physical clock for the quantum subsystem with the kinematical action, showing that this allows to obtain a unitary matter dynamics with quantum-gravity corrections.
		
		We consider a theory consisting of a single scalar matter field $\phi$ with potential $U_m$, immersed in an assigned quantum gravity background, with the addition of the kinematical action. The generalization to the case of $n$ matter fields is straightforward by replacing $\phi$ with $\sum_{a} \phi_{a}$ and inserting the cross-interaction terms into $U_m$.
		
		For the sake of generality, we will consider here the total superspace without assuming specific symmetries of the problem. For this reason, differently from Sec.~\ref{sec:WKB}, the supermomentum contributions of all the components will be present and the corresponding constraints (which were automatically satisfied in the minisuperspace) will be imposed. The total action of the system then reads:
		\begin{equation}
			\begin{split}
				S^{\mathrm{tot}} &= S^{g} + S^{m} + S^{k} = \int \di x^0 \di^3 x \left[ \Pi_{a} \dot{h}^{a} \right.\\
				&\hphantom{=} + p_{\mu} \dot{y}^{\mu} + \pi \dot{\phi} - N \left(\mathcal{H}^{g} + \mathcal{H}^{m} + \mathcal{H}^{k}\right)\\
				&\hphantom{=}\left. - N^i \left(\mathcal{H}^{g}_i + \mathcal{H}^{m}_i + \mathcal{H}^{k}_i \right) \vphantom{\dot{h}^{a}} \right],
			\end{split}
		\end{equation}
		where the supermomentum of the gravity component can be written in the compact form:
		\begin{equation}
		\label{eq:Hi-gravity}
			\mathcal{H}_i^{g} = -2 h_i \D \cdot \Pi \equiv -2 h_{ij} \D_k \Pi^{kj} ,
		\end{equation}
		being $\D$ the three-dimensional covariant derivative on the ADM hypersurfaces; the supermomentum of the matter component is:
		\begin{equation}
		\label{eq:Hi-matter}
			\mathcal{H}_i^{m} = (\partial_i \phi ) \pi ,
		\end{equation}
		and the super-Hamiltonian and supermomentum contributions associated to the kinematical action are those defined in~\eqref{Hkin}.
		
		In the quantization procedure, writing the momentum $p_{\mu}$ in \eqref{eq:newConstrKinAction} as a derivative operator, the total super-Hamiltonian and supermomentum constraints of the system become:
		\begin{subequations}
		\label{eq:TotalConstr}
		\begin{gather}
			(\hat{\mathcal{H}}^{g} + \hat{\mathcal{H}}^{m})\Psi = - \hat{\mathcal{H}}^{k}\Psi \; \rightarrow \; \iu \hbar n^{\mu} \frac{\delta}{\delta y^{\mu}}\Psi ,\\
			(\hat{\mathcal{H}}^{g}_i + \hat{\mathcal{H}}^{m}_i )\Psi= - \hat{\mathcal{H}}^{k}_i \Psi \; \rightarrow\; \iu \hbar b^{\mu}_i \frac{\delta}{\delta y^{\mu}} \Psi .
		\end{gather}
		\end{subequations}
		In this respect, the kinematical action is shown in~\cite {bib:montani-2002} to correspond in the classical limit to a physical fluid, to some extent thought as the ``materialization" of a reference frame. This emerging fluid would suffer the same problem discussed in~\cite{bib:kuchar-1991}, where its emergence is recovered via the reference frame fixing procedure in the gravity-matter action. Here, the kinematical action is, in principle, added to the full quantum system of gravity and matter, but it is regarded as a fast quantum component, on the same footing of the real quantum matter field. Thus, in the present context, the fluid associated to the kinematical action will not appear in the Hamilton-Jacobi equation, not affecting the standard Einsteinian dynamics of the gravitational background.
		
		Following the BO-like approximation as in \eqref{eq:PsiDecomposition}, we write the wave function as:
		\begin{equation}
		\label{eq:Ansatz}
			\Psi ( h_{a}, \phi, y^{\mu} ) = \psi (h_{a}) \chi (\phi, y^{\mu} ; h_{a})
		\end{equation}
		where the slow-varying semiclassical part depends only on the induced 3D metric, while the ``fast" quantum part depends on the matter field and kinematical action and parametrically on the 3D metrics. This separation is justified by considering the different energy scales of the two components, in a case where the scalar fields act as test fields giving negligible contribution to the background and with a fast dynamics that can be computed at nearly fixed values of the 3D metric tensor. As discussed in~\cite{bib:vilenkin-1989}, the matter fields live on an energy scale far from the Planckian one, so that it is reasonable to assume the WDW equation for the background wave function only as in Eq. \eqref{eq:VilBackground}.
		
		We now perform the WKB expansion of the system with respect to the parameter $M$ linked to the Planck mass as in~\cite{bib:kiefer-1991}; in the BO-like approximation, the ratio between the two components of the wave function is:
		\begin{equation}
		\label{eq:BORatio}
			\frac{\hat{H} \chi (\phi, y^{\mu}; h_{a})}{\hat{H}\psi (h_{a})} = \ord{\frac{1}{M}}
		\end{equation}
		which is the analogous of~\eqref{eq:smallness}. We also assume, as in~\cite{bib:vilenkin-1989}, that the fast $\chi$ function has a very small variation with respect to this parameter, expressed by the magnitude of its derivatives with respect to the background variables $h_{a}$:
		\begin{equation}
		\label{eq:derivQuant}
			\frac{\delta}{\delta h_{a}} \chi(\phi, y^{\mu}; h_{a}) \simeq \mathcal{O} \Bigl( \frac{1}{M} \Bigr) .
		\end{equation}
		Following the procedure in Sec.~\ref{ssec:expansionVilKief}, we perform the WKB expansion of the total wave function up to $\ord{1/M}$, which is sufficient to compute the corrections to the functional Schr\"odinger equation arising from the quantum-gravitational background, obtaining:
		\begin{equation}
		\label{eq:MyAnsatz}
			\Psi (h_{a}, \phi, y^{\mu}) = \eu^{ \frac{\iu}{\hbar} \left(M S_0 + P_1 + \frac{1}{M} P_2\right) } \cdot \eu^{\frac{\iu}{\hbar} \left( Q_1 +\frac{1}{M} Q_2\right)} .
		\end{equation}
		
		We stress that, due to the BO-like approximation, the functions $S_0$ and $P_n$ depend only on the three-dimensional metrics $h_{a}$, while the functions $Q_n$ represent the fast component dependent also on the matter and kinematical variables.
		
		The equations to solve are the constraints of the total system and the constraints satisfied by the background wave function $\psi(h_{a})$, which can be written in the form:
		\begin{subequations}
		\begin{gather}
			\left[ -\frac{\hbar^2}{2M} \left( \nabla_g^2 + g \cdot \nabla_g\right) + MV \right] \psi = 0\\
			2 \iu \hbar \, h_{i} \D\cdot \nabla_g \psi = 0\\
			\begin{split}
				&\left[ -\frac{\hbar^2}{2M} \left( \nabla_g^2 + g \cdot \nabla_g \right) + M V \right.\\
				&\qquad\left. \vphantom{\frac{\hbar^2}{2M}} -\hbar^2 \nabla_m^2 + U_m \right] \Psi = \iu \hbar \, n^{\mu} \frac{\delta}{\delta y^{\mu}} \Psi
			\end{split}\\
			\left( 2h_{i}\,\D \cdot \nabla_g - \partial_i \phi \cdot \nabla_m \right) \Psi = \iu\hbar \, b^{\mu}_i \frac{\delta}{\delta y^{\mu}} \Psi .
		\end{gather}
		\end{subequations}
		Here, the additional term
		\begin{equation}
			g \cdot \nabla_g \equiv g_a \frac{\delta}{\delta h_a}
		\end{equation}
		accounts for a generic factor ordering for the derivative operators (see discussion in~\cite{bib:kiefer-2018}); the wave functions $\psi$ and $\Psi$ are of the form~\eqref{eq:Ansatz}, meaning up to order $1/M$ given by~\eqref{eq:MyAnsatz}.

		The first order of expansion is clearly the order $M$; writing explicitly the actions of the gradients on the exponential wave functions, we obtain:
		\begin{subequations}
		\label{eq:OrderM}
		\begin{gather}	
			\frac{1}{2} \nabla_g S_0 \cdot \nabla_g S_0 + V = 0\\
			-2 h_{k} \D \cdot \nabla_g S_0 = 0 .
		\end{gather}
		\end{subequations}
		Here we recover the Hamilton-Jacobi equation for the purely gravitational part of the wave function; hence, the classical limit of gravity is ensured. The real, classical action $S_0$ can be computed from the first equation. The second equation expresses its invariance under 3D diffeomorphisms, due to the hypothesis of the supermomentum constraint for the gravitational part.
		
		The next order of expansion, $M^0$, brings:
		\begin{subequations}
		\begin{gather}
		\label{eq:SuperHGravM0}
			\begin{split}
				&-\frac{\iu \hbar}{2} \nabla^2_g S_0 + \nabla_g S_0 \cdot \nabla_g P_1\\
				&\qquad -\frac{\iu\hbar}{2} g\cdot \nabla_g S_0 = 0
			\end{split}\\
		\label{eq:SuperMGravM0}
			-2h_{k} \D \cdot \nabla_g P_1 = 0\\
		\label{eq:SuperHtotM0}
			\begin{split}
				&-\frac{\iu\hbar}{2} \nabla_g^2 S_0 +\nabla_g S_0 \cdot \nabla_g P_1 -\frac{\iu\hbar}{2} g \cdot \nabla_g S_0\\
				&\qquad + U_m -\iu\hbar \left(\nabla_m^2 Q_1\right)+ \left(\nabla_m Q_1\right)^2\\
				&\qquad = -n^{\mu} \left( \frac{\delta Q_1}{\delta y^{\mu}} \right)
			\end{split}\\
		\label{eq:SuperMtotM0}
			\begin{split}
				&-2h_{k} \D \cdot \nabla_g P_1 +(\partial_i \phi) \left(\nabla_m Q_1\right)\\
				&\qquad= -b^{\mu}_i \left( \frac{\delta Q_1}{\delta y^{\mu} } \right)
			\end{split}
		\end{gather}
		\end{subequations}
		The Eq. \eqref{eq:SuperHGravM0} allows to compute the function $P_1$, which is also invariant under 3D diffeomorphisms due to \eqref{eq:SuperMGravM0}. The wave function at this order can be rewritten as:
		\begin{equation}
		\label{eq:DefF}
			\begin{split}
				\Psi_0 &= f(h_{a}, y^{\mu}, \phi)\\
				&= e^{\frac{\iu}{\hbar} ( M S_0 + P_1 + Q_1)} = D(h) e^{\frac{\iu}{\hbar} Q_1 }
			\end{split}
		\end{equation}
		
		By plugging \eqref{eq:SuperHGravM0} into \eqref{eq:SuperHtotM0}, and using Eq.~\eqref{eq:DefF}, it is possible to rewrite Eq. \eqref{eq:SuperHtotM0} in an interesting form:
		\begin{equation}
		\label{eq:partialSchrSuperH}
			\left(- \hbar^2 \nabla_m^2 + U_m \right) f = \mathcal{H}^m f =\iu \hbar \, n^{\mu} \frac{\delta}{\delta y^{\mu}} f 
		\end{equation}
		where $\mathcal{H}^m$ is the matter super-Hamiltonian. This equation can be combined with the analogous one obtained by plugging \eqref{eq:SuperMGravM0} into \eqref{eq:SuperMtotM0}, that gives:
		\begin{equation}
		\label{eq:partialSchrSuperM}
			- \iu \hbar (\partial_i \phi) \nabla_m f = \mathcal{H}^m_i f = \iu\hbar \, b^{\mu}_i \frac{\delta }{\delta y^{\mu}} f
		\end{equation}
		where $\mathcal{H}^m_i$ is the matter supermomentum. It is now possible to assemble \eqref{eq:partialSchrSuperH} and \eqref{eq:partialSchrSuperM} with the coefficients $N$ and $N^i$, in order to obtain the definition of the deformation vector \eqref{defVector}. Then, by integrating over the ADM hypersurfaces, the derivative operator becomes independent from the spatial coordinates and can be defined as the time derivative:
		\begin{equation}
		\label{eq:SchrodingerF}
			\begin{split}
				\iu\hbar \frac{\delta}{\delta \tau} f &\equiv \iu\hbar \int_{\Sigma} \di^3 x \left(N n^{\mu} + N^i b^{\mu}_i\right) \frac{\delta }{\delta y^{\mu}} f\\
				& = H f = \iu\hbar \int_{\Sigma} \di^3 x N^{\mu} \frac{\delta}{\delta y^{\mu}} f\\
				&= \int_{\Sigma} \di^3 x \left(N \mathcal{H}^m + N^i \mathcal{H}^m_i\right) f 
			\end{split}
		\end{equation}
		Here we have obtained, through the Dirac implementation and using the definition of the deformation vector, a functional Schr\"odinger equation for matter fields, overlapping with standard quantum field theory.
		This equation expresses the quantum dynamics of the matter field immersed in the gravitational background, with a time parameter $\tau$ that clearly describes a nontrivial evolution.
		
		We stress here the difference in the choice of the time coordinate from the proposals \cite{bib:vilenkin-1989} and \cite{bib:kiefer-1991}, since the time is not recovered from the dependence from the ``slow" variables $\nabla_c$ which was shown to be troublesome in Sec.\ref{ssec:OriginNonUn}, but from the kinematical action variables $y^{\mu}$. These variables are present in the definition of the deformation vector, that here has a geometrical connotation, since its values correspond to choices of ADM foliation on the background. The use of its definition~\eqref{defVector} allows to combine and rewrite the momenta $p_{\mu}$ as a single derivative operator, thus constructing the time parameter for the matter subsystem from the kinematical action itself. Nonetheless, the results are formally the same as in \cite{bib:vilenkin-1989} and \cite{bib:kiefer-1991}, since the Schr\"odinger equation is recovered in all cases. The main difference and consequences of this approach will be visible in the next order of expansion.

		The order $M^{-1}$ gives:
		\begin{subequations}
		\begin{gather}
		\label{eq:superHGravM-1}
			\begin{split}
				&-\frac{\iu\hbar}{2} \nabla_g^2 P_1 +\frac{1}{2} \nabla_g P_1 \cdot\nabla_g P_1\\
				&\qquad +\nabla_g S_0 \cdot\nabla_g P_2 -\frac{\iu\hbar}{2} g\cdot \nabla_g P_1 = 0
			\end{split}\\
		\label{eq:superMGravM-1}
			-2 h_{k} \D \cdot \nabla_g P_2 = 0\\
		\label{eq:superHtotM-1}
			\begin{split}
				&-\frac{\iu\hbar}{2} \nabla_g^2 P_1 +\frac{1}{2} \nabla_g P_1 \cdot\nabla_g P_1\\
				&\qquad + \nabla_g S_0 \cdot\nabla_g P_2 -\frac{\iu\hbar}{2} g\cdot \nabla_g P_1\\
				&\qquad - \iu \hbar \nabla_m^2 Q_2+ 2(\nabla_m Q_1) (\nabla_m Q_2)\\
				&\qquad = -n^{\mu} \frac{\delta Q_2}{\delta y^{\mu}}
			\end{split}\\
		\label{eq:superMtotM-1}
			-2 h_{k} \D\cdot \nabla_g P_2 +(\partial_i \phi) \nabla_m Q_2 =- b^{\mu}_i \frac{\delta Q_2}{\delta y^{\mu}}
		\end{gather}
		\end{subequations}
		The first equation allows to compute the function $P_2$, which is invariant under 3D diffeomorphisms by Eq.~\eqref{eq:superMGravM-1}. The total wave function can be written as:
		\begin{equation}
		\label{eq:defTheta}
			\begin{split}
				\Psi_1 &= \Theta(h_{a}, y^{\mu}, \phi)\\
				&=\eu^{\frac{\iu}{\hbar}\left( M S_0 + P_1 + \frac{1}{M} P_2 + Q_1 +\frac{1}{M}Q_2\right)}\\
				&= A(h) f \eu^{\frac{\iu}{\hbar M}Q_2}
			\end{split}
		\end{equation}
		where $f$ is the function satisfying the Schr\"odinger equation~\eqref{eq:SchrodingerF}. Now using Eq.~\eqref{eq:superHGravM-1} together with the results at the previous orders, after some manipulation Eq. \eqref{eq:superHtotM-1} becomes:
		\begin{equation}
		\label{eq:HTheta}
			\iu\hbar n^{\mu} \frac{\delta}{\delta y^{\mu}} \Theta = \mathcal{H}^m \Theta + \left( \nabla_g S_0 \cdot \nabla_g Q_1 \right) \Theta
		\end{equation}
		where we have omitted the term:
		\begin{equation}
			\begin{split}
				&\hbar^2 \left( \frac{1}{\Theta} \nabla_m \Theta -\frac{1}{f} \nabla_m f\right)^2 \Theta\\
				&\qquad = -\frac{1}{M^2} (\nabla_m Q_2)^2 \Theta
			\end{split}
		\end{equation}
		which is of the order $1/M^2$.
		
		It is here evident that the corrections to the Schr\"odinger equation emerge at this order. To recover the total Hamiltonian, the supermomentum constraint must be used; plugging \eqref{eq:superMGravM-1} into \eqref{eq:superMtotM-1} gives:
		\begin{equation}
		\label{eq:HiTheta}
			\iu\hbar b^{\mu}_i \frac{\delta \Theta}{\delta y^{\mu}} = \mathcal{H}^m_i \Theta -2 (h_{k} \D\cdot \nabla_g Q_1) \Theta
		\end{equation}
		and with the linear combination and integration over the hypersurfaces, that reconstruct the total Hamiltonian of the matter field $\mathcal{H}$, we obtain:
		\begin{equation}
		\label{eq:ThetaSigmaQ}
			\begin{split}
				&\iu\hbar \frac{\delta}{\delta \tau} \Theta = H \Theta + \int_{\Sigma} \di^3 x \left( N \nabla_g S_0 \cdot \nabla_g Q_1 \right.\\
				&\left.\quad- 2 N^k h_{k} \D \cdot \nabla_g Q_1\right) \Theta
			\end{split}
		\end{equation}
		Here, we remark that $\Theta$ is the total wave function of the system up to order $1/M$, as defined in \eqref{eq:defTheta}.
		
		We can now further modify this expression to describe the matter field dynamics only.
		
		In fact, even though the WKB approach allows to solve the equations of the constraints order by order, and so the functions $S_0$ and $Q_1$ present here are already defined by the constraints at the previous orders, it is useful to rewrite the equation \eqref{eq:ThetaSigmaQ} such that only the wave function relative to the matter field $\chi (\phi, y^{\mu} ; h_{ij})$ and the purely geometrical functions $S_0, P_n $ appear. This because the explicit forms of $S_0, P_1$ and $P_2$ are defined by the purely gravitational constraints which can be solved separately, obtaining the expressions to substitute in the equation.
		
		However, some attention is required to replace the total wave function $\Theta$ with the matter wave function $\chi$. Since by assumption the functions $S_0, P_n$ do not depend on the variables $y^{\mu}$ nor $\phi$, they can pass through the derivative operators $\nabla_m$ and $\delta/\delta y^{\mu}$ without changing the result. They can also be taken outside the integral $\int_{\Sigma} \di^3 x$, present in the definition of $\mathcal{H}$ and of the time derivative, since they are functionals of the geometries $h_{ij}$ (the supermomentum constraint for the gravitational part at each order assures that these functions are invariant under 3D diffeomorphisms, so they do not depend on the choice of variables $x^i$ but only on the geometries $h_{ij}$).
		
		To rewrite the corrections in the desired form, we can make use of assumption \eqref{eq:derivQuant}. Then, the equation expressing the dynamics of the matter field, including the corrections due to the quantum gravitational background, becomes:
		\begin{equation}
		\label{eqfinale}
			\begin{split}
				&\iu\hbar \frac{\delta}{\delta \tau} \chi = H \chi + \int_{\Sigma} \di^3 x \left[ \vphantom{\left(\frac{1}{\chi}\right)} N \nabla_g S_0 \cdot \left(-\iu\hbar \nabla_g \chi\right) \right.\\
				&\quad\left. -2 N^k h_{k} \D \cdot \left( \frac{1}{\chi}(-\iu\hbar \nabla_g \chi ) \right) \chi \right]
			\end{split}
		\end{equation}
		Thus, at order $1/M$, we have arrived to write down a functional Schr\"odinger equation containing corrections from the quantum nature of the gravitational field.
		
		We observe that the modification in~\eqref{eqfinale} is indeed Hermitian: the function $\nabla_g S_0$ is associated to the solution of the classical Hamilton-Jacobi equation~\eqref{eq:OrderM} and thus must be real (as seen in~\ref{ssec:expansionVilKief} recalling the original assumption in~\cite{bib:kiefer-1991}), while the remaining parts are the conjugate momenta of the gravitational field $-\iu\hbar \nabla_g$. Thus these corrections contribute to a term whose morphology is clearly unitary, overcoming the problems emerging in the previously analyzed works.
		
		We stress that the corrections here computed are of order $1/M$, where $M$ is the appointed parameter of expansion, so they are of low magnitude due to the used approximation and become relevant near the Planckian scale. Further discussion on this result and possible applications are presented in the following section.

\section{Conclusions\label{sec:Conclusions}}
	Let us go through the steps of our analysis. In Sec.~\ref{sec:WKB} the two WKB expansions in $\hbar$ and in the Planck mass proposed in~\cite{bib:vilenkin-1989,bib:kiefer-1991} have been carefully analyzed and compared. We have offered a derivation of both expansions in a formalism that is similar to that adopted in~\cite{bib:kiefer-2018}. By doing so, we have extended the $\hbar$ expansion to arbitrary orders and found quantum gravity corrections to the quantum sector of the theory, starting from the second order in the expansion parameter. This can be seen in the corrected Schr\"odinger equation in curved space-time~\eqref{eq:schrodingerCorrected_Ordh^2}, that is valid both for the $\hbar$ expansion at order $\hbar^2$ and the Planck mass expansion at order $1/M$, revealing that the corrections are of the same kind.
	
	The non-Hermitian nature of the corrections is better highlighted once the derivatives of the wave function in the classical indexes are expressed in terms of time, as reported in Eq.~\eqref{eq:schrodingerCorrected_v2}.
	
	As for the background sector, the $\hbar$ expansion has yielded a HJ equation, corresponding to the Einstein's equations in the presence of a matter source, and the usual equations of a WKB expansion, see Eqs.~\eqref{eq:VilBackground} and~\cite{bib:landau-quantumMechanics}. We have discussed the fact that this feature is not completely shared by the Planck mass expansion, since, even if the background equations have the same form at each order, in this case the classical limit of matter is excluded.
	
	Though the $\hbar$ model needs for the additional hypothesis of smallness of the quantum subsystem to derive the Schr\"odinger equation, it gains in generality: one may think of the more elegant Planck mass expansion as the sub-case of the $\hbar$ case with a purely geometrical background. Moreover, we have stressed that the origin of this difference can be traced in the way the adiabatic separation between slow and fast degrees of freedom is mathematically realized in the two expansions. The Planck mass is the natural adiabatic parameter to split quantum matter from classical geometry and in this sense it does not admit the matter component in the HJ equation. This is acceptable if one is only interested in the recovery of quantum field theory on curved space-time. However, the Planck mass expansion cannot be applied to cosmology without manually rescaling the matter fields with the Planck mass itself, when the theory is applied to inflation, as discussed in the Introduction.
	
	The rest of the paper has been focused on the problem of unitarity breaking at the quantum gravity order. In Sec.~\ref{ssec:non-unit_Plmassexp} we have shown that the solution proposed in~\cite{bib:kiefer-2018} to solve the nonunitarity problem within the framework of the Planck mass expansion is based on very strong hypotheses, and, thus, it solves the problem only for very peculiar models. The central point of the procedure developed in~\cite{bib:kiefer-2018} is in the eigenvalue equations~\eqref{eq:Kief18-Autov}. Passing from the Hamiltonian operators to their eigenvalues allows for the absorption of the non-Hermitian corrections in the background wave function: this has been done through its redefinition contemporaneously with the quantum wave function made in Eqs.~\eqref{eq:redef1} and~\eqref{eq:redef2}. We have argued that the relations~\eqref{eq:Kief18-Autov} cannot hold at the same time, since it is not true, in general, that $\mathrm{H}$ and $H_m$ commute. The reason of this statement is that $\mathrm{H}$ contains the time derivative of the matter Hamiltonian and, in general, $H_m$ and $\dot{H}_m$ do not commute.
	
	As a counterexample to the procedure of ~\cite{bib:kiefer-2018}, we have shown the noncommutation of the matter Hamiltonian with its time derivative for the toy model of inflation described by eqs.~\eqref{eq:toy}. However, our procedure can be applied to all the models where $H_m$ depends on the background variables: indeed, in this case $\dot{H}_m$ contains the time derivatives of such background variables that can be rewritten using their conjugate momenta, so that a natural problem with the commutation of the two operators can arise.
	
	In Sec.~\ref{sec:BOExpansion}, after having reviewed the expansion based on the exact decomposition of the wave function of the universe proposed in~\cite{bib:bertoni-1996}, we have completed the analysis by addressing the two major issues of that study.
	
	On one side, we have restored the gauge invariance of the theory, that was clearly broken by the authors. This has been done in Eq.~\eqref{eq:psi_s}, by defining the background wave function $\psi_s$ corresponding to the purely quantum wave function $\chi_s$ defined in~\cite{bib:bertoni-1996}; as a consequence, the backreaction experiences a two order shift in the expansion parameter from the order of the Hamilton-Jacobi (HJ) equation, where it appeared in~\cite{bib:bertoni-1996}. The first shift is due to the absence of smallness hypothesis of the quantum subsystem, i.e.\ $H_m \sim \hbar$, in~\cite{bib:bertoni-1996}, that would have made the backreaction appear in the continuity equation, at order $\hbar$. However the redefinition of the background wave function has led to a term that exactly compensates the backreaction in the continuity equation, see Eq.~\eqref{eq:ordM0-classical}. Then, we have shown that the first contribution of the backreaction in the background equations appears at the quantum gravity order, accordingly to~\cite{bib:kiefer-2018}. 	
	
	On the other side, we have made explicit the Laplacian operators in the corrected Schr\"odinger equation~\eqref{eq:bertoni-schrodinger} in terms of time derivatives, for the single geometrical variable model, in order to properly check the unitarity of the time evolution at the quantum gravity order. The result of this analysis is contained in Eq.~\eqref{eq:finalSchrBO}, where the analogue of the corrected Schr\"odinger equation obtained in~\cite{bib:kiefer-1991} (see Eq.~\eqref{eq:schrodingerCorrected_Ordh^2}) have been derived in the formalism of~\cite{bib:bertoni-1996}. This equation shows that, once the complete form of the time evolution operator is made explicit, the problem of unitarity breaking at the quantum gravity order affects also the approach~\cite{bib:bertoni-1996}, as well as the others discussed in this paper.
	
	We argued that neither of the proposed solutions for the nonunitarity problem is actually viable, because while in \cite{bib:bertoni-1996} the real meaning of the Laplacian operator in the slow variables is not properly addressed (the time evolution operator is not unitary), in the proposal of \cite{bib:kiefer-2018} the removal of the undesired terms is operated by assumptions which are not valid in general, holding only for special ad-hoc cases.
	
	Summarizing, we have clarified how the problem of a nonunitary evolution, emerging at the second order in the expansion parameter, is independent of the specific nature of such a parameter. This shortcoming of the WKB formulation seems to be an intrinsic feature of the procedure of decomposing the quantum state into a slow-varying and a fast-varying component in order to define time from the slow classical variables, as shown in Sec.~\ref{ssec:OriginNonUn}.
	
	Finally we concentrated on the principal issue of this analysis, corresponding to a new proposal to introduce a proper time evolution in this scheme, avoiding the nonunitarity problem. We have shown in Sec.~\ref{ssec:OriginNonUn} that the origin of the nonunitary terms lies in the definition of time by the semiclassical variables, i.e.\ Eq.~\eqref{eq:time_M}, that has been implemented in the analyzed works. As seen in~\eqref{eq:origin_nonunit}, the construction of such a time parameter leads to a term that is clearly not unitary, thus suggesting a different choice of the evolution parameter. Such a proposal, based on the construction of the physical clock by using the kinematical action~\cite{bib:kuchar-1981}, has been shown to solve the basic difficulties of the previous formulations, namely the appearance of second time derivatives of the quantum wave function which are clearly nonunitary contributions.
	
	The result obtained by introduction of this term~\eqref{kinAction} has led to a satisfactory physical clock, associated in the classical limit to a physical fluid, that can be retained as the materialization of a reference frame. For further discussion on the role of such a fluid in quantum cosmology, see~\cite{bib:kuchar-1991}. The kinematical action has been here added as a fast quantum component, allowing the desired classical limit, i.e.\ the Hamilton-Jacobi equation~\eqref{eq:OrderM} corresponding to the standard Einsteinian dynamics in vacuum of the gravitational background, as viewed in a WKB expansion of the associated vacuum Wheeler-DeWitt equation. This way, we have recovered a standard functional Schr\"odinger equation for the quantum field at the order $M^0$~\eqref{eq:SchrodingerF}, overlapping to standard quantum field theory, and a functional Schr\"odinger equation containing corrections from the quantum nature of the gravitational field at order $1/M$~\eqref{eqfinale}. The associated time parameter, linked to the foliation of the gravitational background and essentially the reference system, has allowed us to overcome the nonunitary problem and it represents a promising construction to the study of quantum gravitational corrections to quantum field theory.
	
	This result offers then a new investigation tool to evaluate the effect of a nonpurely classical gravitational dynamics on the quantum field theory, in the limit of very small energies involved in the quantum dynamics with respect to the Planckian scale, that allows the perturbative approach. A suitable application of this procedure could be in the inflationary sector (as done in~\cite{bib:kiefer-2016} for scalar and tensor perturbations of the inflationary field) or other cosmological cases, where it would be possible to infer the magnitude and consequences of such corrections.
	
\appendix
\section{Reformulation of phase transformations in BO decomposition\label{APP-phase}}
	We here express the phase transformations on the wave functionals performed by~\cite{bib:bertoni-1996} and in Sec.~\ref{sec:BOExpansion} in a more clean and meaningful way.
	
	Going from the initial functions~$ \psi, \chi $ to the final functions~$ \psi_s, \chi_s $ requires two transformations, one that involves~$ A \sim \avg{\nabla_c} $ and one that involves~$ \avg{H_q} \sim \avg{\partial_\tau} $. Hence, the total transformation is given by eqs.~\eqref{eq:bertoni-tildeVars}, \eqref{eq:chi_s} and~\eqref{eq:psi_s}
	\begin{subequations}
	\begin{gather}
		\begin{split}
			\psi &= \eu^{- \frac{\iu}{\hbar} \int A \di c} \widetilde{\psi}\\
			&= \eu^{- \frac{\iu}{\hbar} \int A \di c} \eu^{- \frac{\iu}{\hbar} \int \avg{H_q} \di \tau} \psi_s
		\end{split}\\
		\chi = \eu^{\frac{\iu}{\hbar} \int A \di c} \widetilde{\chi} = \eu^{\frac{\iu}{\hbar} \int A \di c} \eu^{\frac{\iu}{\hbar} \int \avg{H_q} \di \tau} \chi_s ,
	\end{gather}
	\end{subequations}
	where, given the definition of~$ A $, the first phase resembles a Berry phase. Here $ \partial_\tau $ and $ \nabla_c $ are related through the definition of time \eqref{eq:time}. It is argued that such transformations cannot be taken individually, but form a unique transformation on the system~\cite{bib:prasad_datta-1997}. Moreover, we can write the exponent as
	\begin{equation}
		\frac{\iu}{\hbar} \int (A \di c + \avg{H_q} \di \tau) = \int (\avg{\nabla_c} \di c - \avg{\partial_\tau} \di \tau) ,
	\end{equation}
	where we used Eq.~\eqref{eq:bertoni-connect} and~\eqref{eq:schrodinger} (or equivalently Eq.~\eqref{eq:bertoni-schrodinger}, neglecting the fluctuations). If we choose the classical variables to be~$ \lbrace \tau, h_i \rbrace $ from the beginning, where the $ h_i $ are the degrees of freedom orthogonal to time, the last equation reads (a summation on index~$ i $ is implied)
	\begin{equation}
		\begin{split}
			&\int (\avg{\partial_t} \di \tau + \avg{\partial_{h_i}} \di h_i) - \int \avg{\partial_t} \di \tau\\
			&\qquad = \int \avg{\partial_{h_i}} \di h_i .
		\end{split}
	\end{equation}
	Now defining
	\begin{equation}
		A_i = - \iu \hbar \avg{\partial_{h_i}} ,
	\end{equation}
	the full transformation reads
	\begin{gather}
		\psi = \eu^{- \frac{\iu}{\hbar} \int A_i \di h_i} \psi_s\\
		\chi = \eu^{\frac{\iu}{\hbar} \int A_i \di h_i} \chi_s .
	\end{gather}
	An interesting remark on this point is that such transformation is clearly performed on the hyperplane orthogonal to the time coordinate, a feature that was not evident in~\eqref{eq:bertoni-tildeVars}.
	
\section{Detailed calculation of the exact BO decomposition\label{APP-BODecomposition}}
	Working with the assumption of a single gravitational degree of freedom $\alpha$, as stated in Sec.~\ref{ssec:EnhancementBO}, we have
	\begin{subequations}
	\label{eq:alpha_to_t}
	\begin{gather}
		\partial_\tau = \mathcal{G}_{\alpha\alpha} (\partial_\alpha S_0) \partial_\alpha\\
		\partial_\tau S_0 = - 2 V\\
	\label{eq:Gtt}
		\mathcal{G}_{\tau\tau} = - \frac{1}{2 V}\\
	\label{eq:partial_alpha_to_partial_t}
		\begin{split}
			\partial_\alpha &= \frac{\di}{\di \alpha} = \frac{\di S_0}{\di \alpha} \frac{1}{\di S_0 / \di \tau} \frac{\di}{\di \tau}\\
			&= - \frac{1}{2 V} (\partial_\alpha S_0) \partial_\tau .
		\end{split}
	\end{gather}
	\end{subequations}
	We stress that, even with a single geometrical degree of freedom, this procedure is valid only if~$ \spR \neq 0 $, otherwise we would have~$ \partial_\tau S_0 = V = 0 $.

	By using these relations and Eq.~\eqref{eq:schrodinger}, it is easy to find
	\begin{subequations}
	\label{eq:timeDerivatives}
	\begin{gather}
		\iu \hbar \partial_\alpha \widetilde{\chi} = - \frac{1}{2 V} (\partial_\alpha S_0) (H_q - \avg{H_q}) \widetilde{\chi}\\
		\begin{split}
			&\mathcal{G}_{\alpha\alpha} (\iu \hbar \partial_\alpha)^2 \widetilde{\chi} = \left[ \frac{\iu \hbar \dot{V}}{2 V^2} (H_q - \avg{H_q}) \right.\\
			&\quad- \frac{\iu \hbar}{2 V} \mathcal{G}_{\alpha\alpha} (\partial_\alpha^2 S_0) (H_q - \avg{H_q})\\
			&\quad - \frac{\iu \hbar}{2 V} \left( \dot{H}_q - \avg{\dot{H}_q} \right)\\
			&\quad\left. \vphantom{\frac{\iu \hbar \dot{V}}{2 V^2}} - \frac{1}{2 V} (H_q - \avg{H_q})^2 \right] \widetilde{\chi}
		\end{split}\\
	\label{eq:timeDevMean}
		\avg*{\mathcal{G}_{\alpha\alpha} (\iu \hbar \partial_\alpha)^2} = - \frac{1}{2 V} \left( \avg*{H_q^2} - \avg{H_q}^2 \right)
	\end{gather}
	\end{subequations}
	where we expressed time derivatives with a dot and we used the identity
	\begin{equation}
		\partial_\tau \avg{H_q} = \avg{\partial_\tau H_q} ,
	\end{equation}
	due once again to Eq.~\eqref{eq:schrodinger}. Now we can make use of eqs.~\eqref{eq:timeDerivatives}, eqs.~\eqref{eq:background-ord1} and Eq.~\eqref{eq:chi_s}, substituting them in Eq.~\eqref{eq:bertoni_tildeQuantum}; after some cumbersome calculations we obtain the corrected Schr\"odinger equation up to order~$ \ord{1/M} $ as
	\begin{equation}
	\label{eq:finalSchr}
		\begin{split}
			&\iu \hbar \partial_\tau \chi_s = H_q \chi_s - \frac{1}{4 M V} \left[ \left( H_q^2 - \avg*{H_q^2} \right) \vphantom{\frac{\dot{V}}{V}} \right.\\
			&\quad\left. + \iu \hbar (\dot{H}_q - \avg{\dot{H}_q}) - \iu \hbar \frac{\dot{V}}{V} (H_q - \avg{H_q}) \right] \chi_s .
		\end{split}
	\end{equation}
	For completeness, we now turn our attention to the background wave function at order~$ \ord{1/M} $. Rewriting $ \partial_\alpha $ through Eq.~\eqref{eq:partial_alpha_to_partial_t}, with the help of eqs.~\eqref{eq:background-ord1} and of~\eqref{eq:ordM-classical} we find
	\begin{equation}
		\begin{split}
			&- \frac{\hbar^2}{2} \mathcal{G}_{\alpha\alpha} \frac{\partial_\alpha^2 \widetilde{\psi}}{\widetilde{\psi}} = \hbar^2 \partial_\tau \zeta_2 - \iu \hbar^2 \partial_\tau \rho_2\\
			&\quad- \frac{1}{4 V} \avg{H_q}^2 + \frac{\iu \hbar \dot{V}}{4 V^2} \avg{H_q} - \frac{\iu \hbar}{4 V} \avg{\dot{H}_q}\\
			&\quad- \frac{\hbar^2}{4 V} (\partial_\tau \rho_1)^2 - \frac{\hbar^2 \dot{V}}{4 V^2} \partial_\tau \rho_1 + \frac{\hbar^2}{4 V} \partial_\tau^2 \rho_1 .
		\end{split}
	\end{equation}
	Through the last equation and Eq.~\eqref{eq:timeDevMean}, one can rewrite the background equation at the desired order. After separating the real and imaginary parts, we find
	\begin{subequations}
	\label{eq:background-ord1overM}
	\begin{gather}
	\label{eq:background-ord1overM-S}
		\begin{split}
			&\partial_\tau \zeta_2 - \frac{1}{4 V} \left[ (\partial_\tau \rho_1)^2 + \frac{\dot{V}}{V} \partial_\tau \rho_1 - \partial_\tau^2 \rho_1 \right]\\
			&\quad- \frac{1}{4 V} \frac{\avg{H_q^2}}{\hbar^2} = 0
		\end{split}\\
		\partial_\tau \rho_2 - \frac{\dot{V}}{4 V^2} \frac{\avg{H_q}}{\hbar} + \frac{1}{4 V} \frac{\avg{\dot{H}_q}}{\hbar} = 0 .
	\end{gather}
	\end{subequations} 
	This perturbative order clearly shows the backreaction of the quantum subsystem, at the same order expected in~\cite{bib:kiefer-2018}, although the solutions are different, as well as for the corrected Schr\"odinger equation~\eqref{eq:finalSchr}. By writing the equations in the time component from the beginning, Eq.~\eqref{eq:background-ord1-eta} becomes
	\begin{equation}
		\frac{1}{2} \mathcal{G}_{\tau\tau} \partial_\tau^2 S_0 + \partial_\tau \rho_1 = 0 ,
	\end{equation}
	and making use of eqs.~\eqref{eq:alpha_to_t}, we can write
	\begin{equation}
		\partial_\tau \rho_1 = - \frac{\dot{V}}{2 V} .
	\end{equation}
	With this result, we can simplify Eq.~\eqref{eq:background-ord1overM-S} and obtain
	\begin{equation}
		\partial_\tau \zeta_2 - \frac{1}{4 V} \left[ \frac{\ddot{V}}{2 V} - \frac{3}{4} \frac{(\dot{V})^2}{V^2} \right] - \frac{1}{4 V} \frac{\avg{H_m^2}}{\hbar^2} = 0 .
	\end{equation}
	
\bibliographystyle{unsrt}
\bibliography{ms}

\end{document}